\documentclass[12pt,draftclsnofoot,onecolumn]{IEEEtran}

\usepackage{graphicx}
\usepackage{color}
\usepackage{placeins}
\usepackage{float}
\usepackage{tabularx,colortbl}
\usepackage{amssymb}

\usepackage{amsthm}
\usepackage{cite}
\usepackage{amsmath}
\usepackage{caption2}

\usepackage{cases,subeqnarray}
\usepackage{bm,multirow,bigstrut}
\usepackage{textcomp}
\usepackage{latexsym,bm}
\usepackage{booktabs,changebar}
\usepackage{xcolor}
\usepackage{mathtools}
\usepackage{dsfont}
\usepackage{extarrows}
\usepackage{subfigure}

\usepackage{algorithm}
\usepackage{algpseudocode}

\theoremstyle{plain}
\newtheorem{thm}{Theorem}

\newtheorem{rem}{Remark}
\newtheorem{cor}{Corollary}

\IEEEoverridecommandlockouts
\begin{document}

\title{Uplink Performance of RIS-aided Cell-Free Massive MIMO System with Electromagnetic Interference}
\author{Enyu Shi,~\IEEEmembership{Student Member,~IEEE}, Jiayi~Zhang,~\IEEEmembership{Senior Member,~IEEE}, \\Derrick Wing Kwan Ng,~\IEEEmembership{Fellow,~IEEE}, and Bo Ai,~\IEEEmembership{Fellow,~IEEE}

\thanks{E. Shi and J. Zhang are with the School of Electronics and Information Engineering, Beijing Jiaotong University, Beijing 100044, P. R. China. (e-mail: jiayizhang@bjtu.edu.cn).}
\thanks{D. W. K. Ng is with the School of Electrical Engineering and Telecommunications, University of New South Wales, NSW 2052, Australia. (e-mail: w.k.ng@unsw.edu.au).}
\thanks{B. Ai is with the State Key Laboratory of Rail Traffic Control and Safety, Beijing Jiaotong University, Beijing 100044, China. (e-mail: boai@bjtu.edu.cn).}

}

\maketitle
\vspace{-10mm}
\begin{abstract}
Cell-free (CF) massive multiple-input multiple-output (MIMO) and reconfigurable intelligent surface (RIS) are two promising technologies for realizing future beyond-fifth generation (B5G) networks.
In this paper, we consider a practical spatially correlated RIS-aided CF massive MIMO system with multi-antenna access points (APs) over spatially correlated fading channels. Different from previous work, the electromagnetic interference (EMI) at RIS is considered to further characterize the system performance of the actual environment.
Then, we derive the closed-form expression for the system spectral efficiency (SE) with the maximum ratio (MR) combining at the APs and the large-scale fading decoding (LSFD) at the central processing unit (CPU).
Moreover, to counteract the near-far effect and EMI, we propose practical fractional power control (FPC) and max-min power control algorithms to further improve the system performance.
We unveil the impact of EMI, channel correlations, and different signal processing methods on the uplink SE of user equipments (UEs). The accuracy of our derived analytical results is verified by extensive Monte-Carlo simulations.
Our results show that the EMI can substantially degrade the SE, especially for those UEs with unsatisfactory channel conditions. Besides, increasing the number of RIS elements is always beneficial in terms of the SE, but with diminishing returns when the number of RIS elements is sufficiently large. Furthermore, the existence of spatial correlations among RIS elements can deteriorate the system performance when RIS is impaired by EMI.
\end{abstract}

\begin{IEEEkeywords}
Reconfigurable intelligent surface, cell-free massive MIMO, electromagnetic interference, spatial correlation, spectral efficiency.
\end{IEEEkeywords}

\IEEEpeerreviewmaketitle

\section{Introduction}
\IEEEPARstart{T}{he} goal of the fifth-generation (5G) wireless networks is to increase the network capacity by a factor of 1,000 compared to previous generations of networks, providing ubiquitous wireless connectivity to at least 100 billion devices worldwide \cite{saad2019vision}. Recently, with the large-scale commercial deployment of 5G around the world, the global industry has begun to conduct preliminary research on the sixth-generation (6G) of mobile communication technology \cite{giordani2020toward}. In the last few decades, several new technologies have been introduced including millimeter-wave communications \cite{rappaport2017overview}, massive multiple-input multiple-output (MIMO)\cite{larsson2014massive}, and network densification \cite{bjornson2016deploying}. Among them, massive MIMO is the most appealing that has attracted much attention as it can guarantee excellent quality of services to numerous users in the networks. Furthermore, in various scenarios, the net throughput of massive MIMO systems can approach the Shannon capacity via simple linear processing techniques such as zero-forcing (ZF) or maximum ratio (MR) processing \cite{van2020power}. Besides, massive MIMO architectures enjoy the advantage of low-backhaul requirements as the base station (BS) antennas are installed in a form of a compact array. However, traditional cellular networks suffer from severe inter-cell interference. In particular, users at cell borders suffer from both high inter-cell interference and path loss leading to degraded system performance \cite{shi2022spatially}. As such, novel advanced signal processing methods are needed to overcome the inter-cell interference inherent in traditional cellular network deployments.

Cell-free (CF) massive MIMO has recently been introduced as a promising future technology for realizing beyond-5G wireless communication systems \cite{zhang2020prospective}. Indeed, CF massive MIMO is a network architecture consisting of a large number of geographically distributed access points (APs), which are connected to the central processing unit (CPU) and coherently serve all the user equipments (UEs) by performing spatial multiplexing on the same time-frequency resources \cite{ngo2017cell}. Therefore, CF massive MIMO is a major leap of massive MIMO technology to mitigate inter-cell interference, which is the fundamental performance limitation of dense cellular networks \cite{ngo2017cell,zhang2021improving,buzzi2019user}. Recently, a large number of important aspects and fundamentals of CF massive MIMO have been studied to unlock the potential of this architecture. For example, \cite{ngo2017total,du2021cell} studied the system performance that relies on a distributed implementation with the MR processing. Also, the authors in \cite{bjornson2019making} introduced the partially and fully centralized signal processing in CF massive MIMO systems with minimum mean square error (MMSE) combining schemes that can achieve higher spectral efficiency (SE). Besides, the two-layer decoding method for CF massive MIMO is considered an effective decoding scheme with the MR/MMSE combining methods adopted in the first layer decoder and the large-scale fading decoding (LSFD) method applied to the second layer decoder \cite{ozdogan2019performance,wang2022uplink}. Despite the various advantage and potential, Cell-Free massive MIMO still cannot guarantee an appropriate quality of service (QoS) under harsh propagation conditions, such as in the presence of poor scattering environments or severe signal attenuation caused by the presence of large obstacles \cite{van2021reconfigurable}. Meanwhile, with the skyrocketing demand for wireless data throughput, other advanced technologies are also required to satisfy the ever-stringent QoS requirement.

Reconfigurable intelligent surface (RIS) is an emerging technology that is capable of shaping the radio waves at the electromagnetic level without applying sophisticated digital signal processing methods and active power amplifiers \cite{zhang2022reconfigurable}. Specifically, with massive reflective and refractive elements at RIS, effective passive beamforming can be performed by altering the phase of the reflected impinging signals to realize reconfigurable wireless channels/radio propagation environments \cite{wu2019towards,di2020smart,wu2021intelligent,di2020hybrid}. Besides, RIS is fabricated with low-power and low-cost which can be flexibly deployed to assist users with unsatisfactory channel conditions to improve communication quality \cite{yang2020coverage,chen2020resource}.
Due to its unique fabrication, RIS also poses many properties that are different from the traditional uniform linear array (ULA). For example, \cite{bjornson2020rayleigh} proposed a sinc function-based model to capture the spatial correlations that are determined by the spacing between two neighboring RIS elements. Besides, in \cite{de2021electromagnetic}, the authors explored the scenario of having electromagnetic interference (EMI) impinging on RIS and pointed out its significant impact on the system performance. Therefore, in RIS-assisted systems, it is particularly important to consider the existence of EMI and the inherent characteristics of RIS, such as the spatial correlation caused by subwavelength element sizes.

Recently, some works have focused on how to jointly exploit the advantages of CF massive MIMO and RIS to further improve communication performance \cite{9743355,zhang2021joint,ma2022cooperative}. For instance, in \cite{zhang2021joint}, a joint precoding framework for RIS-aided CF massive MIMO systems was proposed to improve the UE communication quality. Besides, in \cite{9743355} the authors studied an RIS-aided CF wireless energy transfer (WET) framework to improve the system capacity and energy efficiency. Furthermore, in \cite{van2021reconfigurable} the author introduced an aggregated channel estimation approach and investigate the RIS-aided CF massive MIMO systems with MR combining over spatially-correlated channels. However, \cite{van2021reconfigurable} considered the Rayleigh fading model of UE-RIS and RIS-AP channels that is not generally valid. In practice, the RIS is generally deployed such that a direct line-of-sight path can be established to obtain better performance gains \cite{wu2019intelligent}. In other words, the Rayleigh fading model adopted in \cite{van2021reconfigurable,9806349} fails to capture this important characteristic. Also, \cite{van2021reconfigurable} adopted single-layer signal processing, which certainly limits the CF massive MIMO system performance.
On the other hand, a common practice in the design of RIS is to only consider the signals generated by the desired system and thereby ignore the EMI or ``noises" that is inevitably presented in the communication environment \cite{middleton1977statistical}. Also, in \cite{chandra2022downlink}, the authors studied the RIS-assisted ultra-reliable low-latency communication (URLLC) in the presence of EMI and the results showed that its existence can reduce the system performance significantly. Besides, the authors considered the physical layer security of RIS-assisted communications and showed the potential of EMI-type interference in degrading the physical layer security in \cite{vega2022physical}. In particular, there is a large amount of EMI in CF massive MIMO systems, which would seriously affect the system performance \cite{larsson2014massive}. In other words, existing theoretical analyzes from the previous research are not applicable to RIS-aided CF massive MIMO systems with EMI. As such, the performance limits of RIS-aided CF massive MIMO systems should be carefully analyzed and designed.

Motivated by the aforementioned observation, we investigate the performance of the RIS-aided CF massive MIMO system over spatially correlated channels with spatial EMI impinging RIS. More specifically, we propose the MMSE-based channel estimation to estimate the aggregated channels between the APs and the UEs. Local processing with MR combining method at the APs and centralized LSFD at CPU are considered in the uplink data transmission. In particular, the uplink performance of the CF massive MIMO system and RIS-aided CF massive MIMO system without EMI are analyzed for comparison. Then, we design two power control optimization algorithms to compensate for the effects of EMI. The specific contributions of the work are listed as follows:
\begin{itemize}
\item We first derive the closed-form expression for the uplink SE of the RIS-aided CF massive MIMO system over spatially correlated channels taking into account spatial EMI. Our results show that the EMI at RIS has a significant impact on the system performance, especially for the UEs with poor channel conditions. Thus, effective interference cancellation schemes, e.g., LSFD cooperation, are critical for improving the system performance. In addition, the RIS-aided CF massive MIMO system achieves higher uplink SE performance than the counterpart without RIS, especially for the edge UEs, the improvement is significant.

\item We analyze the effect of different system parameters and RIS characteristics on the performance, such as the number of AP antennas, RIS elements, spatial correlation, and RIS location. We find that increasing the number of AP antennas magnifies the negative impact of EMI and increasing the number of RIS elements are always beneficial, but with diminishing returns when the number of RIS elements is sufficiently large. Besides, the existence of spatial correlation of RIS elements degrades the system performance, which is even more pronounced when EMI is considered.

\item We design the fractional power control method and the max-min power control algorithm to alleviate the negative impacts of EMI to further improve the system performance for practical implementation. The results show that the proposed power control schemes can further improve the SE of the UEs with poor performance. However, when the EMI power increases, the efficacy of the two power controls decreases. As such, it is particularly important to deploy the RIS in a low EMI environment for RIS-aided CF massive MIMO systems.
\end{itemize}

The rest of this paper is organized as follows. In Section \uppercase\expandafter{\romannumeral2}, we describe the RIS-aided CF massive MIMO system model incorporating the combined effects of spatial correlation, EMI, and channel estimation error. Next, Section \uppercase\expandafter{\romannumeral3} presents the achievable uplink SE of the considered system in the existence of EMI and designs two power control algorithms to compensate the performance degradation caused by the EMI. Then, numerical results and performance analysis are provided in Section \uppercase\expandafter{\romannumeral4}. Finally, Section \uppercase\expandafter{\romannumeral5} concludes this paper.

\textbf{Notation:} The superscripts $\mathbf{x}^{H}$ and $x^\mathrm{*}$ are used to represent conjugate transpose and conjugate, respectively.
The matrices and column vectors are denoted by boldface uppercase letters $\mathbf{X}$ and boldface lowercase letters $\mathbf{x}$, respectively.
The ${\rm{mod}}\left( { \cdot , \cdot } \right)$, $\left\|  \cdot  \right\|$, and $\left\lfloor  \cdot  \right\rfloor $ denote the modulus operation, the Euclidean norm and the truncated argument, respectively. ${\rm{tr}}\left(  \cdot  \right)$, $\mathbb{E}\left\{  \cdot  \right\}$, and ${\rm{Cov}}\left\{  \cdot  \right\}$ are the trace, expectation and covariance operators. $\otimes$ denotes the Kronecker products. We use ${\text{diag}}\left( {{a_1}, \cdots ,{a_n}} \right)$ to denote a block-diagonal matrix with the square matrices ${{a_1}, \cdots ,{a_n}}$ on the diagonal.
Finally, the circularly symmetric complex Gaussian random variable $x$ with variance $\sigma^2$ is denoted by $x \sim \mathcal{C}\mathcal{N}\left( {0,{\sigma^2}} \right)$.

\section{System Model}\label{se:model}
As shown in Figure 1, we consider a RIS-aided CF mMIMO system consisting of $M$ APs, $K$ UEs, one RIS, and a CPU. We consider that each AP is equipped with $L$ antennas and each UE is equipped with a single antenna. All the APs are connected to the CPU via fronthaul links and serve all the UEs via the same time and frequency resources.\footnote{By adopting the central processing, the CPU can utilize the global information from all the involved APs and optimize the signal decoding coefficients with the global information to achieve a better system performance at the expense of higher complexity and more fronthaul signaling overhead \cite{bjornson2019making}.} The RIS has $N$ reflective elements that can introduce some phase shifts to the incident signals \cite{yu2021smart}. Besides, we consider that there exists some EMI at the RIS as EMI inevitably presents in any environment \cite{middleton1977statistical}. Without loss of generality, the standard time division duplex (TDD) protocol is adopted in our system, where the length of each coherence time block is denoted as $\tau_c$. We consider that $\tau_p$ symbols are exploited for the channel estimation phase in the uplink and $\tau_u = \tau_c - \tau_p$ symbols are utilized for data transmission. The channel model, the channel estimation method, and the uplink data transmission phase are then given in the following section.

\begin{figure}[t]
\centering
\includegraphics[scale=0.65]{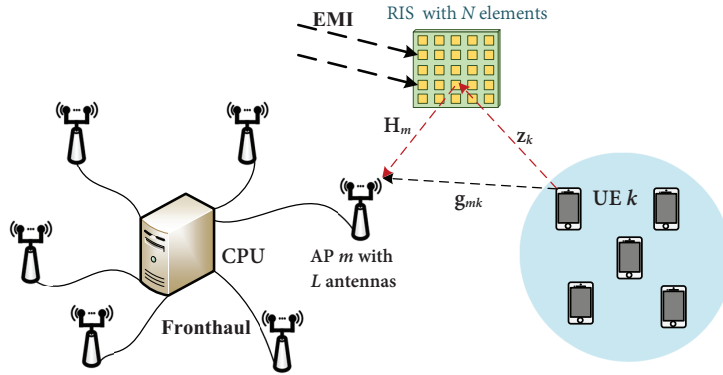}
\caption{An RIS-aided CF mMIMO system with EMI.
\label{Figure_1}}
\end{figure}
\subsection{Channel Model}
We assume a quasi-static block fading model such that in each coherence time block the channels are frequency flat and static. In Figure 1, the channels are divided into two types: the direct link from the UE to the AP and the cascaded link through RIS.
Let ${{\bf{g}}_{mk}}\in {\mathbb{C}}{^{L}}$ denote the direct link channel between UE $k$ and AP $m$. Variable ${{\bf{H}}_m} \in{\mathbb{C}} {^{N \times L}}$ denotes the channel matrix from the RIS to AP $m$ and ${{\bf{z}}_k} \in {\mathbb{C}}{^{N}}$ denotes the channel from UE $k$ to the RIS, respectively. We consider a realistic model to capture the spatial correlation among the RIS scattering elements \cite{bjornson2020rayleigh}, which is due to the sub-wavelength element sizes, sub-wavelength inter-element-distances, and geometric layout. Meanwhile, the spatial correlation at the AP multiple antennas is also considered in this paper. In \cite{van2021reconfigurable}, the authors assume all the channels are Rayleigh fading, while in practice, RIS is usually deployed such that there is a direct path experiencing Rician fading between AP-RIS and UE-RIS for communication enhancement. On the other hand, the channels between the UEs and the APs lack a direct path due to the multi-scatterer distribution in an urban environment. Hence, we focus on a general case that AP-UE channels are Rayleigh fading whereas RIS-UE and AP-RIS channels are Rician fading. Then, the channels ${{\bf{g}}_{mk}}$, ${{\bf{z}}_k}$, and ${{\bf{H}}_m}$ can be modeled as
\begin{align}
{{\bf{g}}_{mk}} &\sim {\cal C}{\cal N}\left( {{\mathbf{0}},{{\mathbf{R}}_{mk}}} \right), \forall m,k,\\
{{\bf{H}}_m} &= {{{\bf{\bar H}}}_m} + {{{\bf{\tilde H}}}_m},\;{{\bf{z}}_k} = {{\bf{\Theta }}_k}{{{\bf{\bar z}}}_k} + {{{\bf{\tilde z}}}_k}, \forall m,k,
\end{align}
where ${\mathbf{R}}_{mk}\in {\mathbb{C}}{^{L \times L}}$ is the spatial correlation matrix of the antennas at AP $m$ and ${\beta _{mk}} = {{{\rm{tr}}\left( {{{\bf{R}}_{mk}}} \right)} \mathord{\left/
 {\vphantom {{{\rm{tr}}\left( {{{\bf{R}}_{mk}}} \right)} L}} \right.
 \kern-\nulldelimiterspace} L}$ is the large-scale fading coefficient between AP $m$ and UE $k$. ${{{\bf{\bar H}}}_m} \in {{\mathbb{C}}^{N \times L}}$ and ${{{\bf{\bar z}}}_k} \in {{\mathbb{C}}^{N}}$ represent the deterministic LoS component of the corresponding AP-RIS and RIS-UE channels, respectively. ${{{\bf{\tilde H}}}_m} \sim {\cal C}{\cal N}\left( {{\bf{0}},{{{\bf{\tilde R}}}_m}} \right)$ and ${{{\bf{\tilde z}}}_k} \sim {\cal C}{\cal N}\left( {{\bf{0}},{{{\bf{\tilde R}}}_k}} \right)$ are the NLoS components, where the spatial covariance matrices ${{\bf{\tilde R}}_m}\in {{\mathbb{C}}^{NL \times NL}}$ and ${{\bf{\tilde R}}_k}\in {{\mathbb{C}}^{N \times N}}$ are given by \cite{wang2022uplink}
\begin{align}
&{{{\bf{\tilde R}}}_m} = \frac{1}{{LN{\beta _m}}}\left( {{\bf{R}}_m^T \otimes {{\bf{R}}_r}} \right), \\
&{{{\bf{\tilde R}}}_k} = {\beta _k}{A_r}{\bf{R}},
\end{align}
respectively, where ${{\beta _m}}$ and ${{\beta _k}}$ denote the large-scale fading coefficients from AP $m$ and UE $k$ to RIS, respectively. Note that ${{\bf{R}}_m}\in {{\mathbb{C}}^{L \times L}}$ and ${{{\bf{R}}_r} = {\beta_m}{A_r}\bf{R}}$ denote the correlation matrices of the AP side and RIS side, respectively. ${A_r} = {d_H}{d_V}$ denotes the area of each RIS element, where ${d_H}$ and ${d_V}$ are the horizontal width and the vertical height of each RIS element, respectively. ${\mathbf{R}} \in {{\mathbb{C}}^{N \times N}}$ denotes the spatial correlation of the RIS which has the $\left( {m',n'} \right)$-th element as ${{\left[ \mathbf{R} \right]_{m'n'}} = {\rm{sinc}}\left( {{{2\left\| {{{\mathbf{u}}_{m'}} - {{\mathbf{u}}_{n'}}} \right\|} \mathord{\left/
 {\vphantom {{2\left\| {{{\mathbf{u}}_{m'}} - {{\mathbf{u}}_{n'}}} \right\|} \lambda }} \right.
 \kern-\nulldelimiterspace} \lambda }} \right)}$, where ${{\rm{sinc}}\left( y \right) = {{\sin \left( {\pi y} \right)} \mathord{\left/
 {\vphantom {{\sin \left( {\pi y} \right)} {\left( {\pi y} \right)}}} \right.
 \kern-\nulldelimiterspace} {\left( {\pi y} \right)}}}$ denotes the sinc function and ${\lambda}$ denotes the carrier wavelength \cite{bjornson2020rayleigh}.
Besides, ${{\mathbf{u}_x} \!=\! {\left[ {0,\bmod \!\left( {x \!-\! 1,{N_H}} \!\right){d_H},\left\lfloor {{{\left( {x \!-\! 1} \right)} \mathord{\left/
 {\vphantom {{\left( {x - 1} \right)} {{N_H}}}} \right.
 \kern-\nulldelimiterspace} {{N_H}}}} \right\rfloor {d_V}} \right]^T}}$\!,\! $x \!\in\! \left\{ {m',n'} \right\}$ is the position vector, where ${N_{V}}$ and ${N_{H}}$ are the numbers of elements at RIS in each column and row, respectively, such that ${N = {N_H} \times {N_V}}$.
Moreover, ${{\bf{\Theta }}_k} = {\rm{diag}}\left( {{e^{j{\theta _{1k}}}}, \cdots ,{e^{j{\theta_{Nk}}}}} \right) \in {{\mathbb{C}}^{N \times N}}$, where ${\theta _{nk}} \in \left[ { - \pi ,\pi } \right]$ is the phase-shift of the LoS component between UE $k$ and the $n$-th element of RIS caused by the small changes in the location of UE $k$ \cite{ozdogan2019performance}. We assume that all the elements of ${{\bf{\Theta }}_k}$ are identical \cite{ozdogan2019performance} such that the LoS component in (2) can be written as ${{{\bf{\bar z}}}_k}{e^{j{\theta _k}}}$.

Let ${\mathbf{\Phi}  = {\rm{diag}}\left( {{e^{j{\varphi _1}}},{e^{j{\varphi _2}}}, \cdots ,{e^{j{\varphi _N}}}} \right)}$ denote the phase shift matrix of the RIS, where ${\varphi _n} \in \left[ { - \pi ,\pi } \right],\forall n \in \left\{ {1, \ldots ,N} \right\}$, denotes the phase shift introduced by the $n$-th RIS element. Thus, the total uplink aggregated channel between AP $m$ and UE $k$ can be formulated as
\begin{align}
{{\bf{o}}_{mk}} = {{\bf{g}}_{mk}} + {\bf{H}}_m^H{\bf{\Phi }}{{\bf{z}}_k},\forall m,k,
\end{align}
which consists of a cascaded link reflected by the RIS and a direct link between AP $m$ and UE $k$. Through simple mathematical derivation, the end-to-end channel between AP $m$ and UE $k$ via the RIS is obtained as
\begin{align}\label{o_mk}
{{\bf{o}}_{mk}} &= {{\bf{g}}_{mk}} + {\bf{H}}_m^H{\bf{\Phi }}{{\bf{z}}_k}\notag\\
 &= {{\bf{g}}_{mk}} + {\left( {{{{\bf{\bar H}}}_m} + {{{\bf{\tilde H}}}_m}} \right)^H}{\bf{\Phi }}\left( {{{{\bf{\bar z}}}_k}{{e^{j{\theta _k}}}} + {{{\bf{\tilde z}}}_k}} \right)\notag\\
 &= \underbrace {{\bf{\bar H}}_m^H{\bf{\Phi }}{{{\bf{\bar z}}}_k}}_{{{{\bf{\bar o}}}_{mk}}}{e^{j{\theta _k}}}  + \underbrace {{{\bf{g}}_{mk}} + {\bf{\bar H}}_m^H{\bf{\Phi }}{{{\bf{\tilde z}}}_k} + {\bf{\tilde H}}_m^H{\bf{\Phi }}{{{\bf{\bar z}}}_k}{e^{j{\theta _k}}} + {\bf{\tilde H}}_m^H{\bf{\Phi }}{{{\bf{\tilde z}}}_k}}_{{{{\bf{\tilde o}}}_{mk}}}.
\end{align}
We assume that UEs move slowly such that the LoS component ${{{{\bf{\bar o}}}_{mk}}}$ is slow time-varying and is known at the APs \cite{ozdogan2019performance}. In the following, we first analyze the second-order statistic of the unknown ${{\bf{\tilde o}}_{mk}}$ that will be handy for the analysis in the sequel.
\begin{thm}
The covariance matrix of the second term ${{{{\bf{\tilde o}}}_{mk}}}$ in \eqref{o_mk} can be obtained as
\begin{align}\label{cov}
\mathbb{E}\left\{ {{ {{{\bf{\tilde o}}}_{mk}}{{{\bf{\tilde o}}}^{H}_{mk}} }} \right\} \!=\! \underbrace {{{\bf{R}}_{mk}} \!+\! {\bf{\bar H}}_m^H\mathbf{\Phi} {{{\bf{\tilde R}}}_k}{\mathbf{\Phi} ^H}{{{\bf{\bar H}}}_m} \!+\! {\bf{Q}}_{mk}^1 \!+\! {\bf{Q}}_{mk}^2}_{{\bf{R}}_{mk}^o}.
\end{align}
\end{thm}
\begin{IEEEproof}
The proof is given in Appendix A.
\end{IEEEproof}
\begin{rem}
Note that since the mean and covariance matrix of the aggregated channel ${{{{\bf{o}}}_{mk}}}$ can be obtained at AP $m$ in each coherence time block, we can utilize them to estimate the end-to-end aggregated channel in the later section.
\end{rem}
\subsection{Electromagnetic Interference Model}
The EMI is produced by a superposition of a continuum of incoming plane waves that are generated by some external sources. In \cite{de2021electromagnetic}, the authors have shown that EMI at RIS has a significant impact on communication performance which cannot be ignored. In particular, the EMI at RIS is modeled as
\begin{align}\label{EMI}
{{\bf{n}}} \sim {\cal C}{\cal N}\left( {{\bf{0}},{A_r}{\sigma_{r} ^2}{\mathbf{R}}} \right),
\end{align}
where ${A_r}$ denote the area of each RIS element and ${{\mathbf{R}}}$ is the spatial correlation matrix of RIS, respectively. ${\sigma_{r}^2}$ is the EMI power at RIS. To proceed further, we define  \cite{de2021electromagnetic}
\begin{align}\label{rho}
\rho  = \frac{{{p_{\max }}\sum\limits_{m = 1}^M {{\beta _m}} }}{{M\sigma _r^2}},
\end{align}
where ${{p_{\max }}}$ denotes the maximum transmit power of the UEs which have the same maximum power. $\rho$ corresponds to the ratio between the received signal power and the EMI power at each antenna element of the RIS, which is to facilitate the description of the EMI power.
\begin{rem}
The correlation of RIS in \cite{bjornson2020rayleigh} shows that the correlation does not exist only when the RIS are positioned along a straight line at a spacing of an integer multiple of $\frac{\lambda }{2}$, which can never happen with a practical two-dimensional RIS. Therefore, it is necessary to consider the effect of EMI with spatial correlation at the RIS elements. Besides, from \eqref{EMI}, we find that both ${A_r}$ and ${\mathbf{R}}$ are affected by the size of each RIS element and the increases of spatial correlation can increase the impact of EMI.
\end{rem}

\subsection{Uplink Pilot Training and Channel Estimation}
We adopt $\tau_p$ pilot sequences which are mutually orthogonal for channel estimation in each coherence time block. All the UEs share the same $\tau_p$ orthogonal uplink pilot sequences \cite{bjornson2019making}. In particular, the pilot sequence of UE $k$ is denoted by ${\bm{\phi} _k} \in {{\mathbb{C}}^{{\tau _p}}}$ and satisfies ${{\left\| {{\bm{\phi} _k}} \right\|^2} = {\tau_p}}$.
Let ${{\cal P}_k}$ denote the index subset of the UEs which adopts the same pilot sequence as UE $k$ including itself as $K > {\tau _p}$. Different from the ideal assumption adopted in \cite{van2021reconfigurable,zhang2021joint,di2020smart}, there are ambient electromagnetic waves in the space and would also be reflected by the RIS and eventually received by the APs \cite{de2021electromagnetic}. Considering the EMI, the received signal ${\bf{X}}_k^p \in {\mathbb{C}^{N \times {\tau _p}}}$ at the RIS is
\begin{align}
{{\mathbf{X}}_k^p} = {{\mathbf{z}}_k}{\bm{\phi}} _k^T + {\mathbf{N}},
\end{align}
where ${\mathbf{N}} \in {\mathbb{C}^{N \times {\tau _p}}}$ denotes the EMI collected over the ${\tau_p}$ samples, produced by the impinging and uncontrollable electromagnetic waves, with ${\mathbf{n}} = {\mathbf{N}}\left( {:,} \right) \sim \mathcal{C}\mathcal{N}\left( {0,{A_r}{\sigma_{r} ^2}{\mathbf{R}}} \right)$. The signal transmitted by UE $k$ is reflected by the RIS and the received signal ${\mathbf{Y}}_{mk}^p \in {\mathbb{C}^{L \times {\tau _p}}}$ at AP ${m}$ can be written as
\begin{align}
{\mathbf{Y}}_{mk}^p &= \sqrt {{{\hat p}_k}} {\mathbf{H}}_m^H{\mathbf{\Phi }}{{\mathbf{X}}_k^p} + \sqrt {{{\hat p}_k}} {{\mathbf{g}}_{mk}}{\bm{\phi}} _k^T + {{\mathbf{N}}_m} \notag\\&= \sqrt {{{\hat p}_k}} \left( {{\mathbf{H}}_m^H{\mathbf{\Phi }}{{\mathbf{z}}_k} + {{\mathbf{g}}_{mk}}} \right){\bm{\phi}} _k^T + {\mathbf{H}}_m^H{\mathbf{\Phi N}} + {{\mathbf{N}}_m}  \notag\\
   &= \sqrt {{{\hat p}_k}} {{\mathbf{o}}_{mk}}{\bm{\phi}} _k^T + {\mathbf{H}}_m^H{\mathbf{\Phi N}} + {{\mathbf{N}}_m} ,
\end{align}
where ${{{\hat p}_k}}$ is the pilot transmit power of UE $k$, ${{\mathbf{N}}_m} \in {\mathbb{C}^{L \times {\tau _p}}}$ is the additive noise with independent ${\cal C}{\cal N}\left( {0,{\sigma ^2}} \right)$ entries, and ${\sigma ^2}$ is the noise power. Then, the signals from all the UEs received at AP $m$ can be written as
\begin{align}
{\mathbf{Y}}_m^p = \sum\limits_{k = 1}^K {\sqrt {{{\hat p}_k}} {{\mathbf{o}}_{mk}}{\bm{\phi}} _k^T + } {\mathbf{H}}_m^H{\mathbf{\Phi N}} + {{\mathbf{N}}_m}.
\end{align}
Then, we multiply the received signal by ${{\bm{\phi}} _k^ *}$ at AP $m$ to estimate ${{{\bf{o}}_{mk}}}$ and the results can be obtained as \cite{wang2022uplink}
\begin{align}\label{y_m}
{\mathbf{y}}_{m}^p &= {\mathbf{Y}}_m^p{\bm{\phi}} _k^ *  = \sum\limits_{k = 1}^K {\sqrt {{{\hat p}_k}} {{\mathbf{o}}_{mk}}{\bm{\phi}} _k^T{\bm{\phi}} _k^ *  + } \left( {{\mathbf{H}}_m^H{\mathbf{\Phi N}} + {{\mathbf{N}}_m}} \right){\bm{\phi}} _k^ *  \hfill \notag\\
&= \sum\limits_{k \in {\mathcal{P}_k}} {\sqrt {{{\hat p}_k}} {\tau _p}{{\mathbf{o}}_{mk}}}  + \left( {{\mathbf{H}}_m^H{\mathbf{\Phi N}} + {{\mathbf{N}}_m}} \right){\bm{\phi}} _k^ *  \hfill \notag\\
&= \sqrt {{{\hat p}_k}} {\tau _p}{{\mathbf{o}}_{mk}} + \sum\limits_{i \in {\mathcal{P}_k}\backslash \left\{ k \right\}} {\sqrt {{{\hat p}_i}} {\tau _p}{{\mathbf{o}}_{mi}}}  + {\mathbf{n}}_{mk}^p,
\end{align}
where ${\mathbf{n}}_{mk}^p = \left( {{\mathbf{H}}_m^H{\mathbf{\Phi N}} + {{\mathbf{N}}_m}} \right){\bm{\phi}} _k^ * $ denotes the total noise at AP $m$. Note that the noise here is different from traditional one as in \cite{9743355,zhang2021joint,ma2022cooperative}. The noise generated by EMI reaches AP $m$ through channel ${\mathbf{H}}_m$, so its covariance matrix is affected by the channel state information (CSI) from RIS to APs. Then, we derive the covariance matrix of the effective noise for later analysis.

\begin{thm}
The covariance matrix of the noise ${\mathbf{n}}_{mk}^p$ at AP $m$  in \eqref{y_m} can be obtained as
\begin{align}
\label{R_mn}
\mathbb{E}\left\{ {{\mathbf{n}}_{mk}^p{{\left( {{\mathbf{n}}_{mk}^p} \right)}^H}} \right\} = {\tau _p}{{\mathbf{R}}_{mm}} + {\tau _p}{\sigma ^2}{{\mathbf{I}}_L}.
\end{align}
\end{thm}
\begin{IEEEproof}
The proof is given in Appendix B.
\end{IEEEproof}
\begin{rem}
Note that the covariance matrix of the EMI ${\mathbf{n}}_{mk}^p$ including two parts: matrix ${{\mathbf{R}}_{mm}}$ and Gaussian noise ${{\mathbf{I}}_L}$. ${{\mathbf{R}}_{mm}}$ is affected by RIS spatial correlation and RIS-AP channels. It is clear that the increase of spatial correlation will lead to the increase of ${{\mathbf{R}}_{mm}}$ and the degradation of performance.
\end{rem}

Based on \eqref{y_m} and \eqref{R_mn}, if ${{{{\bf{\bar o}}}_{mk}}}$, ${\bf{R}}_{mk}^o$, ${\bf{R}}_{mm}$, and ${{\theta _k}}$ are available at AP $m$, we utilize the MMSE channel estimator in \cite{ozdogan2019performance} to estimate the effective channel ${{\bf{o}}_{mk}}$ as
\begin{align}
  {{{\mathbf{\hat o}}}_{mk}}
  &= {\mathbf{\bar H}}_m^H{\mathbf{\Phi }}{{{\mathbf{\bar z}}}_k}{e^{j{\theta _k}}} + \frac{{\sqrt {{{\hat p}_k}} {\mathbf{R}}_{mk}^o}\left( {{\mathbf{y}}_{mk}^p - {\mathbf{\bar y}}_{mk}^p} \right)}{{\sum\limits_{i \in {\mathcal{P}_k}} {{p_i}{\tau _p}{\mathbf{R}}_{mi}^o}  + {{\mathbf{R}}_{mm}} + {\sigma ^2}{{\mathbf{I}}_L}}} \notag \\
   &= {{{\mathbf{\bar o}}}_{mk}}{e^{j{\theta _k}}} + \sqrt {{{\hat p}_k}} {\mathbf{R}}_{mk}^o{\mathbf{\Psi }}_{mk}^{ - 1}\left( {{\mathbf{y}}_{mk}^p - {\mathbf{\bar y}}_{mk}^p} \right),
\end{align}
where ${\bf{\bar y}}_{mk}^p = \sum\nolimits_{i \in {{\cal P}_k}} {\sqrt {{{\hat p}_i}} } {\tau _p}{{{\bf{\bar o}}}_{mi}}{e^{j{\theta _i}}}$ and ${{\bf{\Psi }}_{mk}} = \sum\nolimits_{i \in {{\cal P}_k}} {{{\hat p}_i}{\tau _p}{\bf{R}}_{mi}^o} + {\mathbf{R}}_{mm} + {\sigma ^2}{{\bf{I}}_L}$. The estimated channel ${{{\bf{\hat o}}}_{mk}}$ and the estimation error ${{{\bf{\tilde o}}}_{mk}} = {{\bf{o}}_{mk}} - {{{\bf{\hat o}}}_{mk}}$ are independent random variable with
\begin{align}
&\mathbb{E}\left\{ {{{{\bf{\hat o}}}_{mk}}\left| {{\varphi _k}} \right.} \right\} = {{{\bf{\bar o}}}_{mk}}{e^{j{\varphi _k}}},\:\:{\rm{Cov}}\left\{ {{{{\bf{\hat o}}}_{mk}}\left| {{\varphi _k}} \right.} \right\} = {{\hat p}_k}{\tau _p}{{\bf{\Omega }}_{mk}},\notag\\
&\mathbb{E}\left\{ {{{{\bf{\tilde o}}}_{mk}}} \right\} = {\bf{0}},\quad \quad \quad \quad \;\;\;{\rm{Cov}}\left\{ {{{{\bf{\tilde o}}}_{mk}}} \right\} = {{\bf{C}}_{mk}},
\end{align}
where ${{\bf{\Omega }}_{mk}} = {\bf{R}}_{mk}^o{\bf{\Psi }}_{mk}^{ - 1}{\bf{R}}_{mk}^o$ and ${{\bf{C}}_{mk}} = {\bf{R}}_{mk}^o - {{\hat p}_k}{\tau _p}{\bf{R}}_{mk}^o{\bf{\Psi }}_{mk}^{ - 1}{\bf{R}}_{mk}^o$. \footnote{The MMSE channel estimator can minimize the channel estimation error and provides a satisfactory system performance in CF massive MIMO system in \cite{ozdogan2019performance}. The MMSE channel estimation still performs well when EMI is considered.}

\section{Uplink Data Transmission and Performance Analysis}
The existing CF massive MIMO system is a three-tier structure composed of the CPU, APs, and UEs. The RIS-aided CF mMIMO system introduces an extra RIS layer which adds a cascading link. To the best of the authors' knowledge, recent works for uplink analysis of RIS-aided CF mMIMO system consider the classical AP local processing structure, e.g., \cite{van2021reconfigurable,zhang2021joint,bashar2020performance}, and the EMI at RIS is ignored. In this section, we first consider the impact of EMI on RIS and introduce a process of detecting the uplink transmitted signals by utilizing the channel estimation method introduced in the previous section.
Then, we utilize the AP local MR combining and the LSFD at the CPU and derive an asymptotic closed-form expression of the uplink SE.\footnote{There are many performance indicators that can be adopted to describe the performance of the system, e.g., bit error rate (BER), outage probability (OP), and SE. The BER and OP are mainly used to obtain insights into the reliability of the system. In this paper, we aim to explore the system effectiveness and SE is a suitable metric to capture this important measure.}

\subsection{Uplink Data Transmission}
In the uplink, we consider that the UEs transmit the uplink data to all the APs simultaneously. Due to the existence of EMI, we first obtain the received signal ${\mathbf{x}}_k^u \in {\mathbb{C}^{N}}$ from UE $k$ at RIS as
\begin{align}
{\mathbf{x}}_k^u = {{\mathbf{z}}_k}{s_k} + {\mathbf{n}},
\end{align}
where ${s_k} \sim {\cal C}{\cal N}\left( {0,{p_k}} \right)$ denotes the uplink signal transmitted by UE $k$ with power ${p_k} = \mathbb{E}\{ {{{\left| {{s_k}} \right|}^2}} \}$. ${{\bf{n}}} \sim {\cal C}{\cal N}\left( {{\bf{0}},{A_r}{{\sigma^2_r}}{{\bf{R}}}} \right)$ denotes the additive EMI noise. Then, the received signal ${{\mathbf{y}}_{mk}} \in {\mathbb{C}^L}$ at AP $m$ can be divided into the path reflected from the RIS and the one directly transmitted from UE $k$, which is expressed as
\begin{align}
{{\mathbf{y}}_{mk}} &= {\mathbf{H}}_m^H{\mathbf{\Phi x}}_k^u + {{\mathbf{g}}_{mk}}{s_k} + {{\mathbf{n}}_m} \notag \\
 &= \left( {{\mathbf{H}}_m^H{\mathbf{\Phi }}{{\mathbf{z}}_k} + {{\mathbf{g}}_{mk}}} \right){s_k} + {\mathbf{H}}_m^H{\mathbf{\Phi n}} + {{\mathbf{n}}_m},
\end{align}
where ${{\mathbf{n}}_m} \sim \mathcal{C}\mathcal{N}\left( {0,{\sigma ^2}{{\mathbf{I}}_L}} \right)$ denotes the additive noise. Since there are $K$ UEs, the total received signal ${{\mathbf{y}}_m} \in {\mathbb{C}^L}$ at AP $m$ can be obtained as
\begin{align}
  {{\mathbf{y}}_m} &= \sum\limits_{k = 1}^K {\left( {{\mathbf{H}}_m^H{\mathbf{\Phi }}{{\mathbf{x}}_k} + {{\mathbf{g}}_{mk}}} \right){s_k}}  + {\mathbf{H}}_m^H{\mathbf{\Phi n}} + {{\mathbf{n}}_m} \notag \\
   &= \sum\limits_{k = 1}^K {{{\mathbf{o}}_{mk}}{s_k}}  + {\mathbf{H}}_m^H{\mathbf{\Phi n}} + {{\mathbf{n}}_m}.
\end{align}

We consider that each AP can process the uplink data locally and intermediately with a combining vector \cite{bjornson2019making,wang2020uplink}. We adopt ${{\bf{v}}_{mk}} \in {\mathbb{C}^L} $ denoting the combining vector which is designed by AP $m$ for UE $k$ and then AP $m$ can obtain the local estimate of ${s_k}$ as
\begin{align}\label{s_k}
\setcounter{equation}{19}
{{\tilde s}_{mk}} = {\mathbf{v}}_{mk}^H{{\mathbf{y}}_m} = \sum\limits_{k = 1}^K {{\mathbf{v}}_{mk}^H{{\mathbf{o}}_{mk}}} {s_k} + {\mathbf{v}}_{mk}^H{\mathbf{H}}_m^H{\mathbf{\Phi n}} + {\mathbf{v}}_{mk}^H{{\mathbf{n}}_m}.
\end{align}
Any linear combining vector is available for \eqref{s_k} and the local channel state information (CSI) in AP $m$ can be used to design ${{\bf{v}}_{mk}}$. Here, we consider the MR combining with ${{\bf{v}}_{mk}}= {\bf{\hat o}}_{mk}$ due to its simplicity \cite{wang2020uplink}.
After the MR combining, all the APs convey the local estimates ${{\tilde s}_{mk}}$ in \eqref{s_k} to the CPU. To further mitigate the inter-user interference \cite{bjornson2019making}, ${{\tilde s}_{mk}}$ are linearly weighted with the weight coefficients ${a_{mk}^ * }\in \mathbb{C}$ at the CPU as
\begin{align}\label{s_khat}
{{\hat s}_k} &= \sum\limits_{m = 1}^M {a_{mk}^ * } {{\tilde s}_{mk}} = \sum\limits_{m = 1}^M {a_{mk}^ * } {\mathbf{v}}_{mk}^H{{\mathbf{o}}_{mk}}{s_k} + \sum\limits_{m = 1}^M {a_{mk}^ * } \left( {\sum\limits_{i \ne k}^K {{\mathbf{v}}_{mk}^H{{\mathbf{o}}_{ml}}} {s_i}} \right) \notag\\
&+ \sum\limits_{m = 1}^M {a_{mk}^ * } {\mathbf{v}}_{mk}^H{{\mathbf{n}}_m} + \sum\limits_{m = 1}^M {a_{mk}^ * } {\mathbf{v}}_{mk}^H{\mathbf{H}}_m^H{\mathbf{\Phi n}} \notag \\
&= {\mathbf{a}}_k^H{{\mathbf{u}}_{kk}}{s_k} + \sum\limits_{i \ne k}^K {{\mathbf{a}}_k^H{{\mathbf{u}}_{ki}}{s_i}}  + {{\mathbf{n}}_k} + {{\mathbf{n}}_{\mathrm{EMI},k}},
\end{align}
where ${{\bf{u}}_{ki}} = {[ {{\bf{v}}_{1k}^H{{\bf{o}}_{1i}}, \cdots ,{\bf{v}}_{Mk}^H{{\bf{o}}_{Mi}}} ]^T} \in \mathbb{C}{^M}$ and ${{\bf{a}}_k} = {[ {{a_{1k}}, \cdots ,{a_{Mk}}} ]^T} \in \mathbb{C} {^M}$ denotes the LSFD coefficient vector. The main difference in \eqref{s_khat} compared to \cite{bjornson2019making,zheng2021impact} is that the noise term consists of two parts. ${{\bf{n}}_k} = \sum\nolimits_{m = 1}^M {a_{mk}^ * {\bf{v}}_{mk}^H{{\bf{n}}_m}} $ and ${{\mathbf{n}}_{{\text{EMI}},k}} = \sum\nolimits_{m = 1}^M {a_{mk}^*} {\mathbf{v}}_{mk}^H{\mathbf{H}}_m^H{\mathbf{\Phi n}}$ denote the noise at all the APs and the EMI reflected by the RIS, respectively.

\subsection{Performance Analysis}
In this section, we analyze the uplink performance of the RIS-aided CF massive MIMO system with the MR combining scheme and LSFD, then investigate the impact of spatial correlation and EMI. Based on \eqref{s_khat}, the uplink achievable SE lower bound of UE $k$ can be obtained by utilizing the use-and-then-forget (UatF) bound \cite{bjornson2019making} as follows
\begin{align}\label{SE}
\setcounter{equation}{21}
{\rm{S}}{{\rm{E}}_k} = \frac{{{\tau _u}}}{{{\tau _c}}}{\log _2}\left( {1 + {\gamma _k}} \right),
\end{align}
with the effective signal-to-interference-plus-noise ratio (SINR) ${\gamma _k}$ given by
\begin{align}\label{gamma}
{\gamma _k} \!=\! \frac{{{p_k}{{\left| {{\bf{a}}_k^H\mathbb{E}\left\{ {{{\bf{u}}_{kk}}} \right\}} \right|}^2}}}{{{\bf{a}}_k^H\!\!\left( {\sum\limits_{i = 1}^K {{p_i}{{\bf{T }}_{ki}}} \!-\! {p_k}{ {\mathbb{E}\!\left\{ {{{\bf{u}}_{kk}}} \right\}}{\mathbb{E}\!\left\{ {{{\bf{u}}^{H}_{kk}}} \right\}} } \!+\! {\sigma ^2}{{\bf{D}}_k}} \!+\! {{{\mathbf{U}}^{\mathrm{EMI}}_k}}  \!\right)\!{{\bf{a}}_k}}},
\end{align}
where ${{\bf{T }}_{ki}} = [ \mathbb{E}{\{ {{\bf{v}}_{mk}^H{{\bf{o}}_{mi}}{\bf{o}}_{m'i}^H{{{\bf{v}}}_{m'k}}} \}:\forall m,m'} ] \in {\mathbb{C}^{M \times M}}$ denotes the inter-user interference caused by the pilot contamination and CSI imperfection. The noise at all the APs is obtained as ${{\bf{D}}_k} = {\rm{diag}}( {\mathbb{E}\{ {{{\| {{{{\bf{v}}}_{1k}}} \|}^2}} \}, \cdots ,\mathbb{E}\{ {{{\| {{{{\bf{v}}}_{Mk}}} \|}^2}} \}} ) \in \mathbb{C}{^{M \times M}}$, and the effect of EMI at RIS is denoted as ${{\mathbf{U}}^{\mathrm{EMI}}_k} = {\text{diag}}( {\mathbb{E}\{ {{{\| {{\mathbf{v}}_{1k}^H{\mathbf{H}}_1^H{\mathbf{\Phi n}}} \|}^2}} \}, \cdots \!,\mathbb{E}\{ {{{\| {{\mathbf{v}}_{Mk}^H{\mathbf{H}}_M^H{\mathbf{\Phi n}}} \|}^2}} \}} ) \!\in \!{\mathbb{C}^{M \times M}}$.

Various previous work on RIS-aided CF mMIMO system considered the application of local processing at the APs and a simple centralized decoding at the CPU \cite{van2021reconfigurable,zhang2021joint,9806349} for simplicity. Yet, it greatly limits system performance. Here, we adopt LSFD at the CPU compared with the simple centralized decoding to further improve the system performance and derive the closed-form of uplink SE.

\subsubsection{Large Scale Fading Decoding (LSFD)}
In this case, each AP preprocess its signal by computing local channel estimation of the received data that are then conveyed to the CPU for final decoding. Notice that ${{{\bf{a}}_k}}$ can be optimized to maximize the SE by the CPU, but only channel statistics are available since the CPU does not have knowledge of the channel estimates. In \cite{bjornson2019making}, this approach is known as LSFD in cellular massive MIMO which can be applied here.

\begin{thm}
The closed-form expression for the uplink SE of UE $k$ is given by \eqref{SE}, where the SINR with LSFD is expressed as \eqref{SINR_A} at the top of the next page. Then, the desired signal can be expressed by the following formula
\begin{align}
\setcounter{equation}{24}
&{{\bf{A}}_k} = {\rm{diag}}\left( {{a_{1k}}, \cdots ,{a_{Mk}}} \right) \in {\mathbb{C}^{M \times M}},\\
&{{\bf{Z}}_k} = {\rm{diag}}\left( {{z_{1k}}, \cdots ,{z_{Mk}}} \right) \in {\mathbb{C}^{M \times M}},\\
&{z_{mk}} = {\rm{tr}}\left( {{p_k}{\tau _p}{{\bf{\Omega }}_{mk}} + {{{\bf{\bar o}}}_{mk}}{{\left( {{{{\bf{\bar o}}}_{mk}}} \right)}^H}} \right).
\end{align}
Also, the definition of non-coherent interference ${{{\bf{\xi }}_{ki}}}$ is shown as
\begin{align}
{{\bf{\Xi }}_{ki}} &= {\rm{diag}}\left( {{\xi _{1,ki}}, \cdots ,{\xi _{M,ki}}} \right) \in {\mathbb{C}^{M \times M}},\\
{\xi _{m,ki}} &= {\hat p_k}{\tau _p}{\rm{tr}}\left( {{\bf{R}}_{mi}^o{{\bf{\Omega }}_{mk}}} \right) + {\bf{\bar o}}_{mk}^H{\bf{R}}_{mi}^o{{{\bf{\bar o}}}_{mk}} \notag\\
&+ {\hat p_k}{\tau _p}{\bf{\bar o}}_{mi}^H{{\bf{\Omega }}_{mk}}{{{\bf{\bar o}}}_{mi}} + {\left| {{\bf{\bar o}}_{mk}^H{{{\bf{\bar o}}}_{mi}}} \right|^2}.
\end{align}
The coherent interference ${{{\bf{\Gamma }}_{ki}}}$ is shown as
\begin{align}
&{{\bf{\Gamma }}_{ki}} = {\hat p_k}{\hat p_i}\tau _p^2{\left| {{\rm{tr}}\left( {{{\bf{A}}_k}{{\bf{\Delta }}_{ki}}} \right)} \right|^2},\\
&{{\bf{\Delta }}_{ki}} = {\rm{diag}}\left( {{\varpi _{1,ki}}, \cdots ,{\varpi _{M,ki}}} \right) \in {{\mathbb{C}}^{M \times M}},\\
&{\varpi _{m,ki}}{\rm{ = tr}}\left( {{\bf{R}}_{mi}^o{\bf{\Psi }}_{mk}^{ - 1}{\bf{R}}_{mk}^o} \right).
\end{align}
Finally, the noise at all the APs and the EMI at RIS is shown as
\begin{align}
&{{\bf{J}}_k} = {\rm{diag}}\left( {{{\left\| {{{{\bf{\bar o}}}_{1k}}} \right\|}^2}, \cdots ,{{\left\| {{{{\bf{\bar o}}}_{Mk}}} \right\|}^2}} \right), \\
&{{\mathbf{W}}_k}{{ = \rm{diag}}}\left( {{w_{1k}}, \cdots ,{w_{Mk}}} \right), \\
&{w_{mk}} \!=\! {{{\mathbf{\bar o}}}^H}_{mk}{{\mathbf{R}}_{mm}}{{{\mathbf{\bar o}}}_{mk}} \!+\! {{\hat p}_k}\tau _p^2{\rm{tr}}\!\left( \!{{{\left( {{\mathbf{\Psi }}_{mk}^{ - 1}} \!\right)}^H}\!{{\left( {{\mathbf{R}}_{mk}^o} \right)}^H}{{\mathbf{R}}_{mm}}{\mathbf{R}}_{mk}^o} \!\right).
\end{align}
\end{thm}

\newcounter{mytempeqncnt}
\begin{figure*}[t!]
\normalsize
\setcounter{mytempeqncnt}{1}
\setcounter{equation}{23}
\begin{align}\label{SINR_A}
\gamma _k^{{\mathrm{EMI}}^{(1)}} = \frac{{{p_k}{{\left| {{\text{tr}}\left( {{\mathbf{A}}_k^H{{\mathbf{Z}}_k}} \right)} \right|}^2}}}{{\sum\limits_{i = 1}^K {{p_i}{\text{tr}}\left( {{\mathbf{A}}_k^H{{\mathbf{\Xi }}_{ki}}{{\mathbf{A}}_k}} \right)}  + \sum\limits_{i \in {\mathcal{P}_k}\backslash \left\{ k \right\}} {{p_i}{{\mathbf{\Gamma }}_{ki}}}  + {\text{tr}}\left( {{\mathbf{A}}_k^H{{\mathbf{W}}_k}{{\mathbf{A}}_k}} \right) + {\text{tr}}\left( {{\mathbf{A}}_k^H\left( {{\sigma ^2}{{\mathbf{Z}}_k} - {p_k}{\mathbf{J}}_k^2} \right){{\mathbf{A}}_k}} \right)}}.
\end{align}
\setcounter{equation}{24}
\hrulefill
\end{figure*}

\begin{IEEEproof}
The proof is given in Appendix C.
\end{IEEEproof}

Furthermore, to maximize the effective SINR in \eqref{SINR_A}, the CPU can utilize LSFD to optimize ${{{\bf{a}}_k}}$ as
\begin{align}\label{ak}
\setcounter{equation}{35}
  {{\mathbf{a}}_k} &= \left( {\sum\limits_{i = 1}^K {{p_i}{{\mathbf{T}}_{ki}}}  - {p_k}\mathbb{E}\left\{ {{{\mathbf{u}}_{kk}}} \right\}\mathbb{E}\left\{ {{\mathbf{u}}_{kk}^H} \right\} + {\sigma ^2}{{\mathbf{D}}_k}} \right. \notag \\
  &{\left. { + {\mathbf{U}}_k^{{\text{EMI}}}} \right)^{ - 1}}\mathbb{E}\left\{ {{{\mathbf{u}}_{kk}}} \right\}.
\end{align}
Then, the SINR in \eqref{gamma} can be derived as
\begin{align}\label{gamma1}
  &{\gamma _k}^{{\text{EM}}{{\text{I}}^{(1)}}} = {p_k}\mathbb{E}\left\{ {{\mathbf{u}}_{kk}^H} \right\} \times \left( {\sum\limits_{i = 1}^K {{p_i}{{\mathbf{T}}_{ki}}} } \right. \notag \\
  &{\left. { - {p_k}\mathbb{E}\left\{ {{{\mathbf{u}}_{kk}}} \right\}\mathbb{E}\left\{ {{\mathbf{u}}_{kk}^H} \right\} + {\sigma ^2}{{\mathbf{D}}_k} + {\mathbf{U}}_k^{{\text{EMI}}}} \right)^{ - 1}}\mathbb{E}\left\{ {{{\mathbf{u}}_{kk}}} \right\} .
\end{align}

\begin{IEEEproof}
The proof of \eqref{ak} and \eqref{gamma1} can be easily derived following from the standard results of matrix derivation in \cite{ozdogan2019performance} and is therefore omitted.
\end{IEEEproof}

\subsubsection{Simple Centralized Decoding}
In this case, the received signals at each AP were processed with MR combining, then sent to the CPU where the signal from different APs was processed with simple centralized decoding \cite{bjornson2019making,wang2022uplink}. Actually, it is just a special case of LSFD and the coefficient is written as ${{\mathbf{a}}_k} = {\left[ {1, \cdots ,1} \right]^T} \in {\mathbb{C}^{M \times 1}}$.
\begin{cor}
For adopting a simple centralized decoding at the CPU, the closed-form expression for the uplink SE of UE $k$ is given by \eqref{SE}, where the SINR is expressed as \eqref{SINR_woA} at the top of this page. For comparison, we also provide a closed-form expression \eqref{SINR_woEMI} without EMI at the top of this page.
\end{cor}

\begin{IEEEproof}
The proof is given in Appendix C.
\end{IEEEproof}

\begin{rem}
Note that the EMI and the spatial correlation via the RIS affect the system performance via ${{\bf{R}}_{mm}}$ expressed in a closed-form in \eqref{SINR_A}. By reducing the spatial correlation of RIS elements or the power of EMI, the reduction of the norm of ${{\bf{R}}_{mm}}$ leads to the reduction of the denominator in \eqref{SINR_A}, and the system performance is improved. Besides, reducing the spatial correlation of RIS elements can increase ${{\bf{R}}_{mk}^o}$, which leads to the increase of the numerator in \eqref{SINR_A}, also, the system performance is improved.
\end{rem}

\begin{rem}
In CF mMIMO systems, the spatial correlation of the AP antennas is beneficial to the SE \cite{wang2020uplink}, while in our system, the aggregated channel and the EMI consists the interaction between ${{\bf{R}}_{mk}}$ and ${{\bf{R}}}$. As such the conclusion in \cite{wang2020uplink} may no longer hold, especially when the RIS elements are strongly correlated.
\end{rem}

\begin{figure*}[t!]
\normalsize
\setcounter{mytempeqncnt}{1}
\setcounter{equation}{37}
\begin{align}\label{SINR_woA}
\gamma _k^{{\mathrm{EMI}}^{(2)}} = \frac{{{p_k}{{\left| {{\text{tr}}\left( {{{\mathbf{Z}}_k}} \right)} \right|}^2}}}{{\sum\limits_{i = 1}^K {{p_i}{\text{tr}}\left( {{{\mathbf{\Xi }}_{ki}}} \right)}  + \sum\limits_{i \in {\mathcal{P}_k}\backslash \left\{ k \right\}} {{p_i}{\hat p_k}{\hat p_i}\tau _p^2{{\left| {{\text{tr}}\left( {{{\mathbf{\Delta }}_{ki}}} \right)} \right|}^2}}  + {\text{tr}}\left( {{{\mathbf{W}}_k}} \right) + {\text{tr}}\left( {{\sigma ^2}{{\mathbf{Z}}_k} - {p_k}{\mathbf{J}}_k^2} \right)}}.
\end{align}
\setcounter{equation}{38}
\hrulefill
\end{figure*}

\begin{figure*}[t!]
\normalsize
\setcounter{mytempeqncnt}{1}
\setcounter{equation}{38}
\begin{align}\label{SINR_woEMI}
\!\!\!\!\!\!\!\!\!\!\!\!\!\!\!\!{\gamma _k} = \frac{{{p_k}{{\left| {{\rm{tr}}\left( {{\bf{A}}_k^H{{\bf{Z}}_k}} \right)} \right|}^2}}}{{\sum\limits_{i = 1}^K {{p_i}{\rm{tr}}\left( {{\bf{A}}_k^H{{\bf{\Xi }}_{ki}}{{\bf{A}}_k}} \right)}  + \sum\limits_{i \in {{\cal P}_k}\backslash \left\{ k \right\}} {{p_i}{{\bf{\Gamma }}_{ki}}}  + {\rm{tr}}\left( {{\bf{A}}_k^H\left( {{\sigma ^2}{{\bf{Z}}_k} - {p_k}{\bf{J}}_k^2} \right){{\bf{A}}_k}} \right)}}.
\end{align}
\setcounter{equation}{39}
\hrulefill
\end{figure*}

\subsection{Power Control Schemes}
The EMI affects the performance of UEs differently in the RIS-aided CF network because of the different locations of the UEs. Therefore, the impacts of EMI can be reduced by the design of power allocation, exploiting the different propagation conditions of the UEs.
\subsubsection{Fractional Power Control}
In general, the system performance is limited by the near-far effects. Transmitting signals at full power not only wastes energy resources, but also causes severe inter-user interference. To address this issue, we extend the FPC scheme proposed in \cite{nikbakht2019uplink} to the considered RIS-aided CF massive MIMO system. Specifically, the FPC scheme for the RIS-aided CF massive MIMO system based on the equivalent large-scale fading coefficients of the considered UE, which captures the average signal strength from the UE to all the APs. Inspired by this, we design the power control factor of UE $k$ as
\begin{align}\label{eta}
{\eta _k} = {\left( {\frac{{\mathop {\min }\limits_k \left( {\sum\nolimits_{m = 1}^M {{\text{tr}}\left( {{\mathbf{R}}_{mk}^o} \right)} } \right)}}{{\sum\nolimits_{m = 1}^M {{\text{tr}}\left( {{\mathbf{R}}_{mk}^o} \right)} }}} \right)^\alpha },\forall k,
\end{align}
where ${{\mathbf{R}}_{mk}^o}$ is given in \eqref{cov} and ${0 < \alpha  < 1}$ denotes the fractional power control parameter. Therefore, the transmission power ${p_k}$ of UE $k$ can be expressed as ${p_k} = {\eta _k}{p_{\max }}$. For the uplink transmission, the fairness of UEs is important and \eqref{eta} aims to improve the performance of UEs with poor connection strength.

\begin{algorithm}[t]
\caption{Bisection Algorithm for Solving \eqref{maxmize} } 
\begin{algorithmic}[1]
\State \textsl{Initialization:} Choose the initial values of ${t_{\min}}$ and ${t_{\max}}$, where ${t_{\min}}$ and ${t_{\max}}$ define a range of relevant values of the objective function in \eqref{maxmize}. Choose a tolerance $\varepsilon  > 0$.      
\State Set $t: = \frac{{{t_{\min }} + {t_{\max }}}}{2}$. Solve the following convex feasibility program:
\begin{align}
\label{suanfa}
\setcounter{equation}{42}
\left\{ {\begin{array}{*{20}{c}}
  {\frac{1}{t}{p_k}{{\left| {{\text{tr}}\left( {{\mathbf{A}}_k^H{{\mathbf{Z}}_k}} \right)} \right|}^2} \geqslant {\zeta _k},\;k = 1, \ldots K,} \\
  {\;0 \leqslant {p_k} \leqslant {p_{\max }},\;k = 1, \ldots K,}
\end{array}} \right.
\end{align}
where ${\zeta _k} = \sum\limits_{i = 1}^K {{p_i}{\text{tr}}\left( {{\mathbf{A}}_k^H{{\mathbf{\Xi }}_{ki}}{{\mathbf{A}}_k}} \right)}  + \sum\limits_{i \in {\mathcal{P}_k}\backslash \left\{ k \right\}} {{p_i}{{\mathbf{\Gamma }}_{ki}}}  + {\text{tr}}\left( {{\mathbf{A}}_k^H{{\mathbf{W}}_k}{{\mathbf{A}}_k}} \right) + {\text{tr}}\left( {{\mathbf{A}}_k^H\left( {{\sigma ^2}{{\mathbf{Z}}_k} - {p_k}{\mathbf{J}}_k^2} \right){{\mathbf{A}}_k}} \right)$.
\State If problem \eqref{suanfa} is feasible, then set ${t_{\min }}: = t$, else set ${t_{\max }}: = t$.
\State Stop if ${t_{\max }} - {t_{\min }} < \varepsilon $. Otherwise, go to step 2.
\end{algorithmic}
\end{algorithm}

\subsubsection{Max-min SE Power Control}
\begin{figure}[t]
\centering
\includegraphics[scale=0.55]{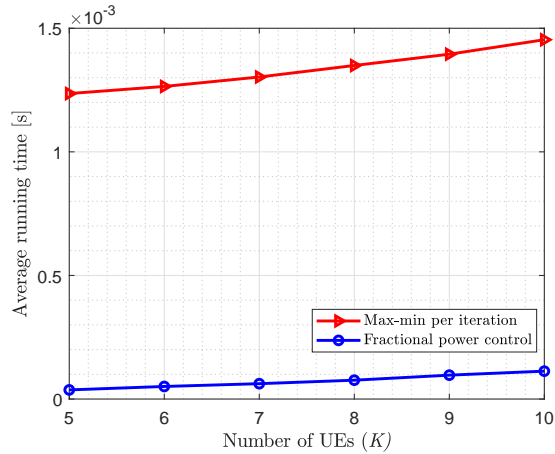}
\caption{Average running time against different number of UEs with different power control methods (${M = 10}$, ${L = 1}$, ${N = 16}$, ${\tau_p = 3}$, ${{d_{\rm{V}}} = {d_{\rm{H}}} = \frac{1}{2}\lambda }$).
\label{Fig_1}}
\end{figure}

To further improve the system performance of poor users, we adopt the max-min SE power control and the optimization problem can be fromulated as
\begin{align}\label{maxmin}
\setcounter{equation}{40}
  &\mathop {{\text{maxmize}}}\limits_{{p_k}} \; \mathop {{\text{min}}}\limits_{k = 1, \cdots ,K} \; \gamma _k^{{\mathrm{EMI}}^{\left( 1 \right)}} \notag \\
  &{\text{subject}}\;{\text{to}}\quad 0 \leqslant {p_k} \leqslant {p_{\max }},\; k = 1, \ldots K,
\end{align}
where $\gamma _k^{{\mathrm{EMI}}^{\left( 1 \right)}}$ is given by \eqref{SINR_A}. Problem \eqref{maxmin} can be equivalently reformulated as
\begin{align}
\label{maxmize}
\setcounter{equation}{41}
  &\mathop {{\text{maxmize}}}\limits_{{p_k},t} \quad t \notag \\
  &{\text{subject}}\;{\text{to}}\quad t \leqslant \gamma _k^{{\mathrm{EMI}}^{\left( 1 \right)}},\;k = 1, \ldots K, \notag \\
  &\quad \quad \quad \quad \;\;\;0 \leqslant {p_k} \leqslant {p_{\max }},\;k = 1, \ldots K,
\end{align}
with $t$ is an auxiliary optimization variable. Consequently, problem \eqref{maxmize} can be efficiently solved by using bisection and solving a sequence of linear feasibility problems as shown in Algorithm 1. Note that in Algorithm 1, if we focus on a reasonable range of the initial upper and lower bounds (i.e., $t_{\min} = 0$ and ${t_{\max }} \geqslant \max \left[ {{\gamma _k}} \right]$), we can effectively find a feasible solution within a few iterations.
\begin{rem}
The computational complexity of Algorithm 1 is $\mathcal{O}\left( {2K^2+2K} \right)$ and it is clear that complexity increases rapidly as the number of users increases. In contrast, the FPC enjoys a linear complexity but at the expense of some performance loss.
We can find from Figure~\ref{Fig_1} that the running times of the two methods increase with the increases of $K$, while that of the max-min method scales with $K$ much faster than that of FPC.
As such, in practice, when the number of UEs is small, exploiting the max-min method is recommended. However, when the number of UEs is large or the communication reliability requirement is not stringent, FPC serves as a better choice.
\end{rem}

\section{Numerical Results and Discussion}
In this section, we provide some numerical results to verify the accuracy of the derived analysis and evaluate the performance of the RIS-aided CF mMIMO system over EMI and the spatial correlation of RIS elements. We assume that the APs and UEs are uniformly distributed in the $0.1 \times 0.1\,{\rm{k}}{{\rm{m}}^2}$ area with a wrap-around scheme \cite{bjornson2019making}, respectively. Moreover, the RIS is located at the regional center. The height of each AP, UE, and RIS is $15$ m, $1.65$ m, and $30$ m, respectively. We set the carrier frequency as 1.9 GHz. Besides, each coherence block consists of ${\tau _c} = 200$ corresponding to a coherence bandwidth of 200 kHz and ${\tau_p} = 3$ are reserved for pilot transmission. For the path loss, we take AP-RIS as an example which consists of an LoS path and we utilize the COST 321 Walfish-Ikegami model \cite{wang2022uplink} to compute the path loss as
\begin{align}
\setcounter{equation}{43}
{\beta _m}\left[ {{\rm{dB}}} \right] =  - 30.18 - 26{\log _{10}}\left( {\frac{{{d_m}}}{{1\:{\rm{m}}}}} \right) + {F_m},
\end{align}
where $d_m$ denotes the distance between AP $m$ and RIS. The Rician $\kappa$-factor is denoted as ${\kappa _m} = {10^{1.3 - 0.003{d_m}}}$.
The shadow fading $F_{m}$ and other parameters is similar to \cite{ngo2017cell} with ${F_m} = \sqrt {{\delta _f}} {a_m} + \sqrt {1 - {\delta _f}} {b_{{\text{RIS}}}}$, where ${a_m} \sim {\cal N}\left( {0,\delta _{{\text{sf}}}^2} \right)$ and ${b_{\rm{RIS}}} \sim {\cal N}\left( {0,\delta _{{\text{sf}}}^2} \right)$ are independent random variables and ${{\delta _f}}$ is the shadow fading parameter. The variances of $a_m$ and $b_{\rm{RIS}}$ are $\mathbb{E}\left\{ {{a_m}{a_{m'}}} \right\} = {2^{ - \frac{{{d_{mm'}}}}{{{d_{{\text{dc}}}}}}}}$, $\mathbb{E}\left\{ {{b_{{\text{RIS}}}}{b_{{\text{RIS'}}}}} \right\} = {2^{ - \frac{{{d_{{\text{RISRIS}}'}}}}{{{d_{{\text{dc}}}}}}}}$, respectively, where ${{d_{mm'}}}$ are the geographical distances between AP $m$-AP $m'$ and ${{d_{{\text{RISRIS}}'}}} = 0$, respectively, $d_{\rm{dc}}$ is the decorrelation distance depending on the environment. We set ${{\delta _f}} = 0.5$, $d_{\rm{dc}} = 100$ m and ${\delta _{{\text{sf}}}} = 8$ in this paper. The large-scale coefficients of ${{\bf{H}}_m}$ are given by
\begin{align}
\beta _m^{{\rm{LoS}}} = \frac{{{\kappa _m}}}{{{\kappa _m} + 1}}{\beta _m},{\kern 1pt} \;\;\beta _m^{{\rm{NLoS}}} = \frac{1}{{{\kappa _m} + 1}}{\beta _m}.
\end{align}
Each AP is equipped with a uniform linear array (ULA) with omnidirectional antennas so the $n$-th element of the deterministic LoS component ${{{\bf{\bar H}}}_m} \in {\mathbb{C}^{N \times L}}$ can be written as ${\left[ {{{{\bf{\bar H}}}_m}} \right]_{n,l}} = \sqrt {\beta _m^{{\rm{LoS}}}} {e^{j2\pi {d_H}\left( {n - 1} \right)\sin \left( {{\theta _m}} \right)}}$, where ${\theta_m}$ is the angle of arrival from AP $m$ to the RIS and ${d_H}$ denotes the antenna spacing parameter (in fractions of the wavelength).
\begin{figure}[t]
\centering
\includegraphics[scale=0.55]{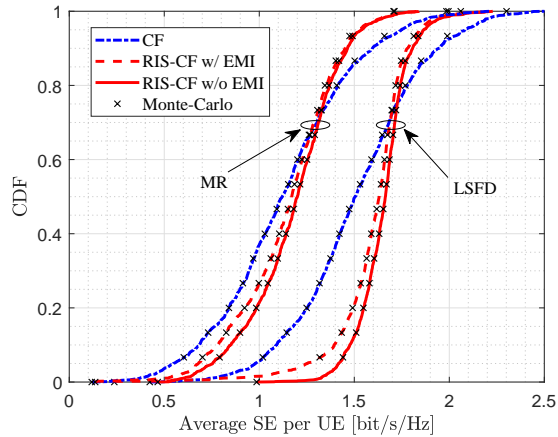}
\caption{CDF of the uplink average SE against different systems over the MR and LSFD combining methods with EMI (${M = 10}$, ${K = 5}$, ${L = 1}$, ${N = 16}$, ${\tau_p = 3}$, ${{d_{\rm{V}}}={d_{\rm{H}}} = \frac{1}{2}\lambda }$, $\rho = 20\:\rm{dB}$).}\vspace{-4mm}
\label{Fig_2}
\end{figure}

\begin{figure}[t]
\centering
\includegraphics[scale=0.55]{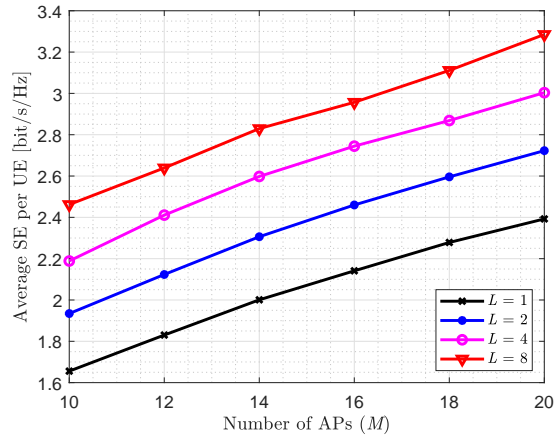}
\caption{Uplink average SE against the number of AP $M$ over different numbers of AP antennas with LSFD ($K = 5$, $N = 16$, $\tau_p = 3$, ${{d_{\rm{V}}}={d_{\rm{H}}} = \frac{1}{2}\lambda }$, $\rho = 20\:\rm{dB}$).}\vspace{-4mm}
\label{Fig_3}
\end{figure}
The Gaussian local scattering model in \cite{9837438} is utilized to generate the spatial correlation matrix ${{\bf{R}}_{mk}}$. The ${{\bf{R}}_{mk}}$ $\left( {l,n} \right)$-th element can be written as
\begin{align}
{\left[ {{{\bf{R}}_{mk}}} \right]_{ln}} = \frac{{\beta _{mk}^{{\rm{NLoS}}}}}{{\sqrt {2\pi } {\sigma _\varphi }}}\int_{ - \infty }^{ + \infty } {{e^{j2\pi {d_H}\left( {l - n} \right)\sin \left( {{\theta _{mk}} + \delta } \right)}}} {e^{ - \frac{{{\delta ^2}}}{{2\sigma _\varphi ^2}}}}d\delta ,
\end{align}
where $\delta  \sim {\cal N}\left( {0,\sigma _\varphi ^2} \right)$ is the distributed deviation from ${\theta_{mk}}$ with angular standard deviation (ASD) ${{\sigma _\varphi }}$. Every UE transmits with a power of $23$ dBm and the noise power  ${\sigma ^2} = 94$ dBm. The fractional power control parameter $\alpha = 0.6$. As for the phase shift design of RIS, we take a fixed value that the $N$ elements phase shift is set equal to ${\pi  \mathord{\left/
 {\vphantom {\pi  4}} \right.
 \kern-\nulldelimiterspace} 4}$ \cite{van2021reconfigurable}.\footnote{We use UatF bound to derive closed solutions. Since we calculate the expected value as shown in \eqref{gamma}, the phase shift effect of RIS is not significant, especially when the spatial correlation is weak.} Note that only Figures~\ref{Fig_8} and~\ref{Fig_9} consider the power control schemes.

Figure~\ref{Fig_2} compares the CDF of the uplink average SE for RIS-aided CF and CF massive MIMO systems under the random locations of UEs and APs with MR/LSFD combining methods with EMI. We illustrate the CDF of the uplink SE by using Monte-Carlo simulations and the proposed analytical framework. It is clear that the RIS-aided CF system has a significant gain over the conventional CF system. Meanwhile, the LSFD achieves a 1.86 times improvement at 95\%-likely points compared with MR combining. Besides, from the 95\%-likely points, we know that the EMI harms the system performance, especially for the UEs with poor performance while its impact on those UEs with high SE is minimal.

\begin{figure}[t]
\centering
\includegraphics[scale=0.55]{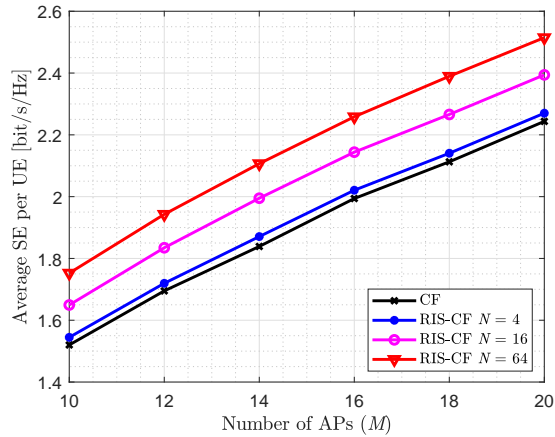}
\caption{Uplink average SE for the CF and RIS-aided CF mMIMO systems against the number of AP $M$ with different numbers of RIS elements with LSFD ($K=5$, $L=1$, $\tau_p = 3$, ${{d_{\rm{V}}}={d_{\rm{H}}} = \frac{1}{2}\lambda }$, $\rho = 20\:\rm{dB}$).}\vspace{-4mm}
\label{Fig_4}
\end{figure}

\begin{figure}[t]
\centering
\includegraphics[scale=0.55]{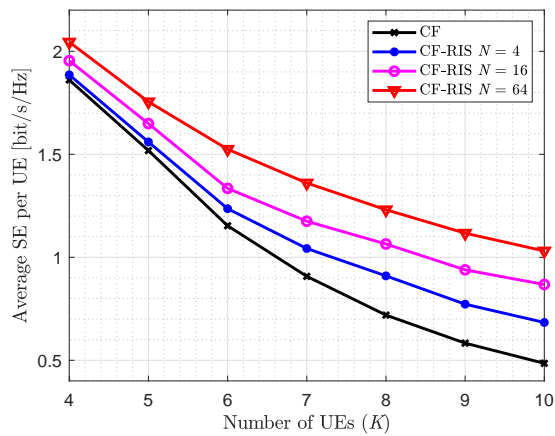}
\caption{Uplink average SE for the CF and RIS-aided CF mMIMO systems against the number of UE $K$ with different numbers of RIS elements with LSFD ($M=10$, $L=1$, $\tau_p = 3$, ${{d_{\rm{V}}}={d_{\rm{H}}} = \frac{1}{2}\lambda }$, $\rho = 20\:\rm{dB}$).}\vspace{-4mm}
\label{Fig_5}
\end{figure}

Figure~\ref{Fig_3} shows the average SE per UE as a function of the number of AP $M$ with the different numbers of per AP's antennas $L$ with LSFD. It is clear that increasing the numbers of AP $M$ and AP antennas $L$ can improve the system performance as the increased number of spatial degrees of freedom facilitates more efficient beamforming. Moreover, it is interesting to find that when $M = 20$, the average SE of $L=2$ achieves a 12.5\% gain compared to that of $L=1$. Yet, for $L=8$, only 10\% of the gain is achieved compared to that of $L=4$, which is much smaller. This reveals that the potential gains due to the AP antennas are diminishing when the number of AP antennas is sufficiently large.

\begin{figure}[t]
\centering
\includegraphics[scale=0.55]{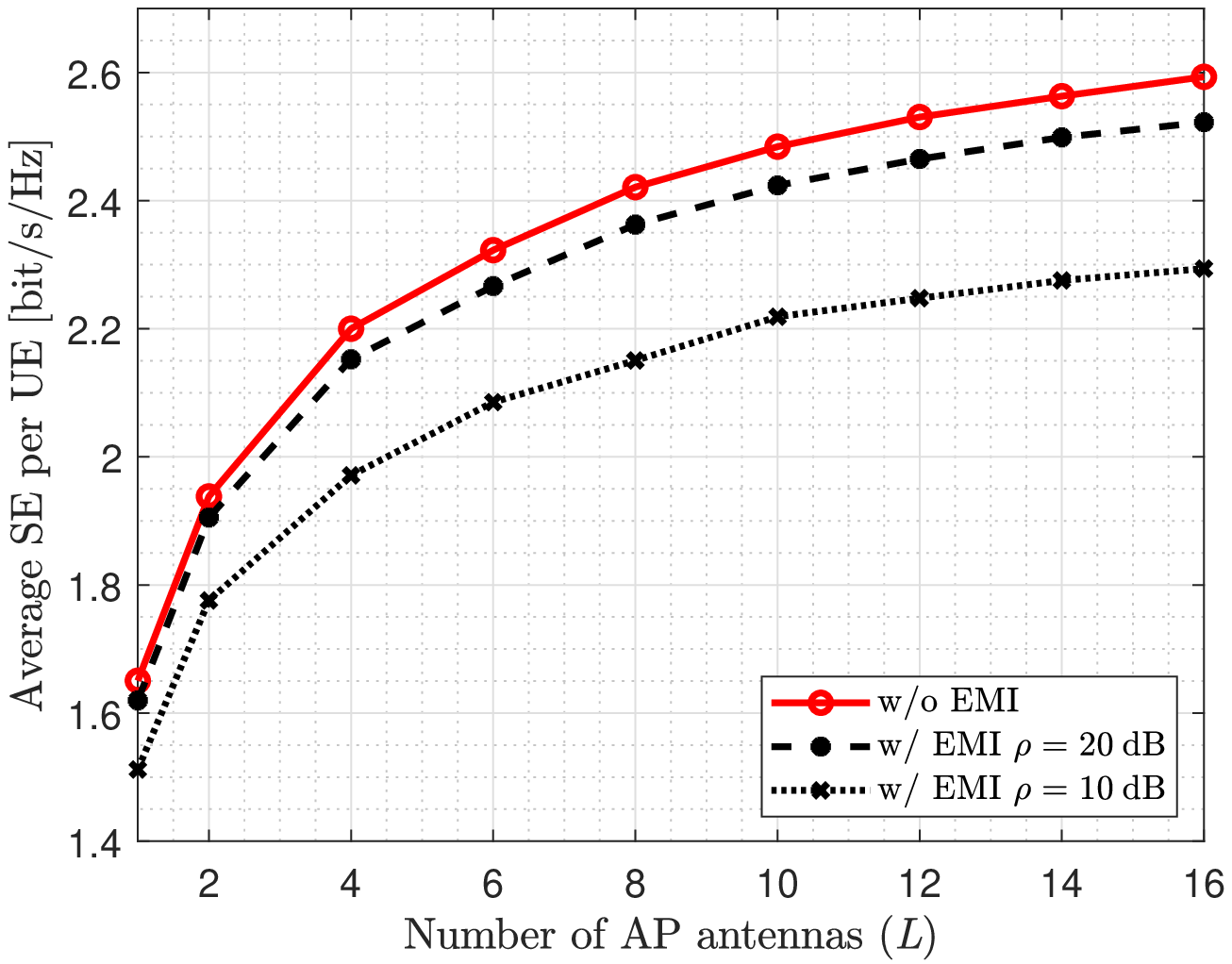}
\caption{Uplink average SE against the number of AP antennas $L$ with different EMI powers with LSFD ($M=10$, $K=5$, $N=16$, $\tau_p = 3$, ${{d_{\rm{V}}}={d_{\rm{H}}} = \frac{1}{2}\lambda }$).} \vspace{-4mm}
\label{Fig_6}
\end{figure}

\begin{figure}[t]
\centering
\includegraphics[scale=0.55]{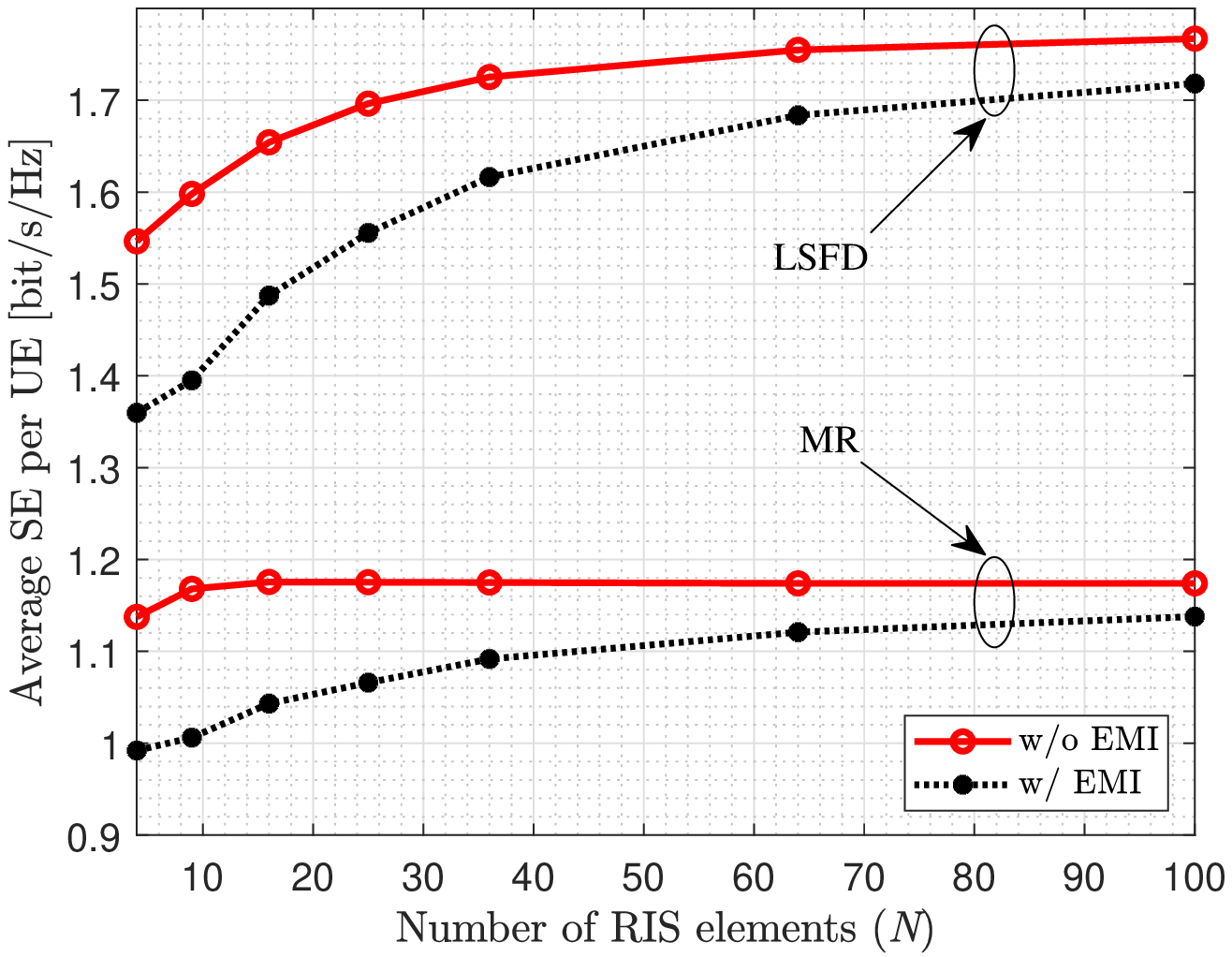}
\caption{Uplink average SE against the number of RIS elements $N$ with MR combining and LSFD ($M=10$, $L=1$, $K=5$, $\tau_p = 3$, ${{d_{\rm{V}}}={d_{\rm{H}}} = \frac{1}{2}\lambda }$, $\rho = 20\:\rm{dB}$).} \vspace{-4mm}
\label{Fig_7}
\end{figure}

Figure~\ref{Fig_4} and Figure~\ref{Fig_5} show the average SE per UE as a function of the number of AP $M$ and UE $K$ with different RIS elements $N$ with LSFD. It is clear that increasing the number of RIS elements $N$ can improve the system performance as it increases the amount of reflected energy in the reflected link. Figure 5 shows that as the number of UEs increases, the system performance decreases, which is caused by the increased interference among the UEs. Note that the performance degradation due to the increased number of UEs can be relieved by increasing the number of RIS elements. In fact, passive beamforming can achieve better interference management. This reveals the importance of deploying RIS in hotspot areas.

Figure~\ref{Fig_6} shows the average SE as a function of the number of AP antennas $L$ with different EMI powers with LSFD. It is clear that the performance gap is enhanced with the increase of $L$ since an increasing number of AP antennas will cause the increased EMI power from RIS. In particular, when the number of AP antennas $L$ is sufficiently large, for example, compared with $L = 14$, $L = 16$ only offers a marginal performance gain. It reveals that continuously increasing the number of AP antennas in a CF system to achieve further system performance improvements is not cost-effective unless more efficient combining methods are employed at the AP rather than the MR combining.

\begin{figure}[t]
\centering
\includegraphics[scale=0.55]{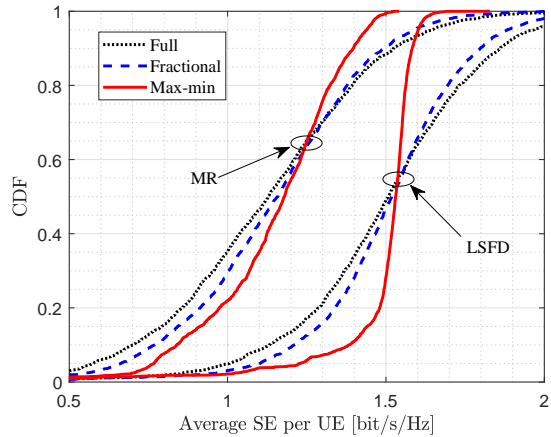}
\caption{CDF of the uplink average SE for the RIS-aided CF massive system over the MR combining and the LSFD with EMI (${M = 10}$, ${K = 5}$, ${L = 1}$, ${N = 16}$, ${\tau_p = 3}$, ${{d_{\rm{V}}}={d_{\rm{H}}} = \frac{1}{2}\lambda }$, $\rho = 20\:\rm{dB}$).} \vspace{-4mm}
\label{Fig_8}
\end{figure}

\begin{figure}[t]
\centering
\includegraphics[scale=0.55]{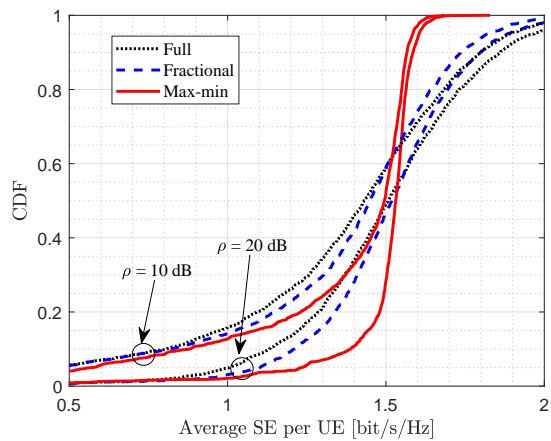}
\caption{CDF of the uplink average SE for the RIS-aided CF massive system against different EMI power with the LSFD (${M = 10}$, ${K = 5}$, ${L = 1}$, ${N = 16}$, ${\tau_p = 3}$, ${{d_{\rm{V}}}={d_{\rm{H}}} = \frac{1}{2}\lambda }$).} \vspace{-4mm}
\label{Fig_9}
\end{figure}

Figure~\ref{Fig_7} shows the average SE as a function of the number of RIS elements $N$ with MR combining and LSFD. It is clear that the performance gap becomes smaller and is less sensitive to the increases of $N$. It reveals that increasing the element number of RIS in RIS-aided CF massive MIMO systems is beneficial to naturalizing the impairment caused by EMI. Moreover, compared with the MR combining, when $N$ is larger than 16, the system performance barely increases and the LSFD behaves similarly after $N$ is larger than 64. This reveals that with a large number of RIS elements, the LSFD can still exploit the spatial DoF offered by the RIS in CF massive MIMO system.

\begin{figure}[t]
\centering
\includegraphics[scale=0.4]{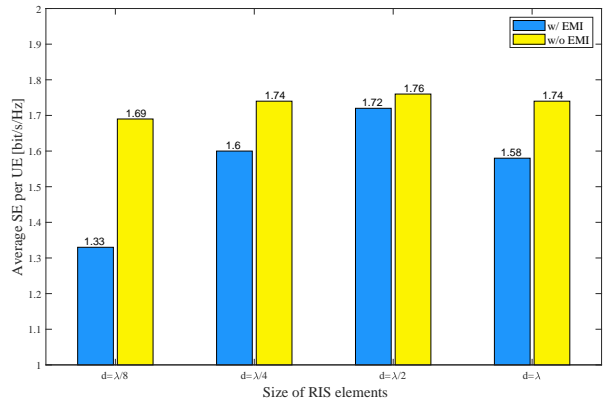}
\caption{Average SE against different sizes of RIS element with and without EMI over the LSFD (${M = 10}$, ${L = 1}$, ${K = 5}$, ${N = 64}$, ${\tau_p = 3}$, $\rho = 20\:\rm{dB}$).\vspace{-4mm}
\label{Fig_10}}
\end{figure}

\begin{figure}[t]
\centering
\includegraphics[scale=0.55]{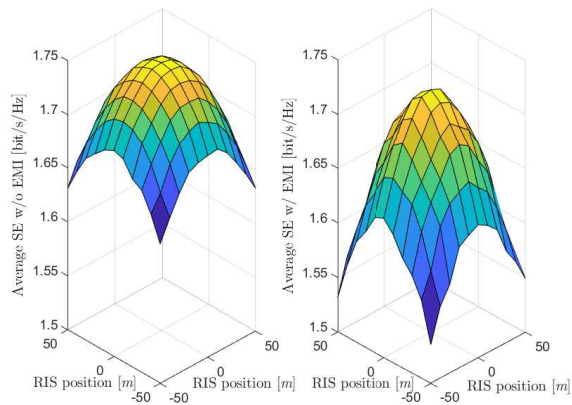}
\caption{Average SE against different RIS positions with and without EMI over the LSFD (${M = 10}$, ${L = 1}$, ${K = 5}$, ${N = 16}$, ${\tau_p = 3}$, ${{d_{\rm{V}}}={d_{\rm{H}}} = \frac{1}{2}\lambda }$).\vspace{-4mm}
\label{Fig_11}}
\end{figure}

Figure~\ref{Fig_8} compares the CDF of the uplink average SE for RIS-aided CF massive MIMO system with MR combining and the LSFD under different power control schemes. It is clear that compared with the proposed fractional power control and the max-min SE power control, we can significantly improve the SE for the UEs with poor performance. Moreover, jointly with the max-min power control, the LSFD and MR combining achieve 25.1\% and 20\% gains compared with the full power method at 90\%-likely points, respectively. Hence, the max-min SE power control with the addition of excellent signal processing methods can further achieve user fairness in the RIS-aided CF massive MIMO system.

Figure~\ref{Fig_9} compares the CDF of the uplink average SE for the RIS-aided CF massive MIMO system against different EMI powers with LSFD. It is clear that the same as in Figure~\ref{Fig_2}, the EMI has a severer negative impact on the UEs with poor performance. In particular, when the EMI is serious, i.e., $\rho = 10\: \rm{dB}$, the performance degradation of poor users is significant even though the max-min SE power control is applied. This result suggests that when the EMI is strong, it is not a sensible option for RIS to continue serving all UEs.

Figure~\ref{Fig_10} shows the effect of the spatial correlation brought by the size of the RIS elements on the average SE of the RIS-aided CF massive MIMO system. It is clear that for the cases with or without EMI, the system performance is the best when the element size is of a half wavelength. Moreover, when EMI is considered, the spatial correlation of RIS has a severer impact on system performance. This can be explained by (8) that the power distribution of EMI is affected by the spatial correlation of RIS. Therefore, the element size of RIS needs to be carefully designed in practice to reduce the impact of spatial correlation and EMI on the system performance.

Figure~\ref{Fig_11} shows the impact of different RIS deployment locations on the system performance with and without EMI over LSFD. It is clear that deploying RIS at the center of the service area can obtain satisfactory system performance regardless of the presence of EMI. In the absence of EMI, the system performance in the region is almost centrosymmetric. However, when there is EMI at RIS, the degradation of the system performance at the edge is magnified, which reveals the importance of deploying RIS at the center of the service area.

\section{Conclusions}
In this paper, we investigated the uplink SE of RIS-aided CF massive MIMO systems over spatially correlated Rician channels with EMI.
Furthermore, we analyzed the uplink SE with the MR combining at the APs and the LSFD at the CPU and obtained the analytical closed-form expression with EMI for characterizing the performance.
It is important that the EMI degrades the performance of the considered systems, especially for the UEs with unsatisfactory channel conditions, while the LSFD cooperation is less sensitive to the EMI compared with MR combining. As such, it is particularly important to deploy the RIS in a low EMI environment for RIS-aided CF massive MIMO systems.
In addition, increasing the number of APs and RIS elements significantly improve the average SE, with diminishing returns when the number of AP antennas and RIS elements is sufficiently large.
Besides, increasing the number of AP antennas also increases the negative impact of EMI. Also, increasing the number of RIS elements is always beneficial as the impairment caused by EMI can be relieved by the extra spatial DoF brought by the increasing number of RIS elements.
Meanwhile, considering the existence of EMI, the spatial correlations among RIS elements impose a severer negative impact on the system performance.
Also, we find that designing the element spacing of RIS as half wavelength can suppress the spatial correlation and EMI to the greatest extent.
Finally, for practical application in RIS-aided CF scenarios, we proposed the practical fractional power control in the considered systems to improve the SE performance of the worst AP and max-min SE power control scheme for preliminary EMI improvement. In future work, we will consider the EMI-aware beamforming design to enable the implementation of RIS-aided CF mMIMO networks.

\begin{appendices}

\section{Proof of Theorem 1}
This appendix calculates the covariance matrix in (5). To start with, we express
\begin{align}
\setcounter{equation}{46}
&\mathbb{E}\left\{ {{ {{{\bf{\tilde o}}}_{mk}}{{{\bf{\tilde o}}}^{H}_{mk}} }} \right\} = {{\bf{R}}_{mk}} + {\bf{\bar H}}_m^H{\bf{\Phi}} {{{\bf{\tilde R}}}_k}{{\bf{\Phi}} ^H}{{{\bf{\bar H}}}_m} \notag\\
 &\!+\! \underbrace {\mathbb{E}\!\left\{\! {{\bf{\tilde H}}_m^H{\bf{\Phi}} {{{\bf{\bar z}}}_k}{\bf{\bar z}}_k^H{{\bf{\Phi}} ^H}{{{\bf{\tilde H}}}_m}} \!\right\}}_{{\bf{Q}}^1_{mk}} \!+\! \underbrace {\mathbb{E}\!\left\{\! {{\bf{\tilde H}}_m^H{\bf{\Phi}} {{{\bf{\tilde z}}}_k}{\bf{\tilde z}}_k^H{{\bf{\Phi }} ^H}{{{\bf{\tilde H}}}_m}} \!\right\}}_{{\bf{Q}}^2_{mk}}\!.
\end{align}
To calculate ${{\bf{Q}}^1_{mk}}$, we let ${{\bf{B}}_k} = {\bf{\Phi }}{{{\bf{\bar z}}}_k}{\bf{\bar z}}_k^H{{\bf{\Phi }}^H}$, and ${{\bf{Q}}^1_{mk}}$ is derived as
\begin{align}
{{\bf{Q}}^1_{mk}} &= \mathbb{E}\left\{ {{\bf{\tilde H}}_m^H{{\bf{B}}_k}{{{\bf{\tilde H}}}_m}} \right\} = \mathbb{E}\left\{ {\begin{array}{*{20}{c}}
{{\bf{\tilde H}}_{1m}^H{{\bf{B}}_k}{{{\bf{\tilde H}}}_{1m}}}& \cdots &{{\bf{\tilde H}}_{1m}^H{{\bf{B}}_k}{{{\bf{\tilde H}}}_{Lm}}}\\
 \vdots & \ddots & \vdots \\
{{\bf{\tilde H}}_{Lm}^H{{\bf{B}}_k}{{{\bf{\tilde H}}}_{1m}}}& \cdots &{{\bf{\tilde H}}_{Lm}^H{{\bf{B}}_k}{{{\bf{\tilde H}}}_{Lm}}}
\end{array}} \right\}.
\end{align}
For each element in ${{\bf{Q}}^1_{mk}}$, we can obtain $\mathbb{E}\!\left\{\! {{\bf{\tilde H}}_{lm}^H{{\bf{B}}_k}{{{\bf{\tilde H}}}_{l'm}}} \!\right\} \! {\buildrel (a) \over =} {\rm{tr}}\!\left( \!\!{{{\bf{B}}_k}{{\left[ {{{{\bf{\tilde R}}}_m}} \right]}_{\left( {lN - N + 1 \sim lN,l'N - N + 1 \sim l'N} \right)}}} \!\right)$, where (a) follows by applying the trace of product property, ${\rm{tr}}\left( {{\bf{XY}}} \right) = {\rm{tr}}\left( {{\bf{YX}}} \right)$, for some given size-matched matrices ${\bf{X}}$ and ${\bf{Y}}$.
We calculate ${{\bf{Q}}^2_{mk}}$ by applying the same method as above, then $Q_{mk}^2$ can be obtained as \eqref{Q2mk} at the top of this page, where the $ll'$-th element is obtained as ${\left[ {{\bf{Q}}_{mk}^2} \right]_{ll'}} = {\rm{tr}}\left( {{\bf{\Phi }}{{{\bf{\tilde R}}}_k}{{\bf{\Phi }}^H}{{\left[ {{{{\bf{\tilde R}}}_m}} \right]}_{\left( {lN - N + 1 \sim lN,l'N - N + 1 \sim l'N} \right)}}} \right)$.

\begin{align}\label{Q2mk}
Q_{mk}^2 \!=\!\!\! \left[ {\begin{array}{*{20}{c}}
  {{\text{tr}}\left( {{\mathbf{\Phi }}{{{\mathbf{\tilde R}}}_k}{{\mathbf{\Phi }}^H}{{\left[ {{{\mathbf{\tilde R}}_m}} \right]}_{\left( {1 \sim n,1 \sim n} \right)}}} \right)}& \cdots &{{\text{tr}}\left( {{\mathbf{\Phi }}{{{\mathbf{\tilde R}}}_k}{{\mathbf{\Phi }}^H}{{\left[ {{{\mathbf{\tilde R}}_m}} \right]}_{\left( {1 \sim n,LN - N + 1 \sim LN} \right)}}} \right)} \\
   \vdots & \ddots & \vdots \\
  \!\!\!{{\text{tr}}\!\left(\!\! {{\mathbf{\Phi }}{{{\mathbf{\tilde R}}}_k}{{\mathbf{\Phi }}^H}\!{{\left[ {{{\mathbf{\tilde R}}_m}} \right]}_{\left( {LN - N + 1 \sim LN,1 \sim n} \right)}}} \right)}& \cdots &{{\text{tr}}\!\left(\!\! {{\mathbf{\Phi }}{{{\mathbf{\tilde R}}}_k}{{\mathbf{\Phi }}^H}\!{{\left[ {{{\mathbf{\tilde R}}_m}} \right]}_{\left( {LN \!-\! N \!+\! 1 \sim LN,LN \!-\! N \!+\! 1 \sim LN} \right)}}} \right)}
\end{array}} \!\!\right],
\end{align}

\section{Proof of Theorem 2}\label{nn}
The covariance matrix in \eqref{R_mn} is calculated in this appendix. First, we express
\begin{align}
\setcounter{equation}{49}
{\mathbf{n}}_{mk}^p = \left( {{\mathbf{H}}_m^H{\mathbf{\Phi N}} + {{\mathbf{N}}_m}} \right){\bm{\phi}} _k^ *  = {\mathbf{H}}_m^H{\mathbf{\Phi N}}{\bm{\phi}} _k^ *  + {{\mathbf{N}}_m}{\bm{\phi}} _k^ *.
\end{align}
Since the noise at different pilot symbol instants is independent and uncorrelated, the covariance matrix of ${\mathbf{n}}_{mk}^p$ can be written as
\begin{align}
\mathbb{E}\!\left\{ {{\mathbf{n}}_{mk}^p{{\left( {{\mathbf{n}}_{mk}^p} \right)}^H}} \!\right\} \!=\! {\tau _p}\mathbb{E}\!\left\{ {{\mathbf{H}}_m^H{\mathbf{\Phi n}}{{\mathbf{n}}^H}{{\mathbf{\Phi }}^H}{{\mathbf{H}}_m}} \!\right\} \!+\! {\tau _p}\mathbb{E}\!\left\{ {{{\mathbf{n}}_m}{\mathbf{n}}_m^H} \right\}\!,
\end{align}
where ${{\bf{n}}} \sim {\cal C}{\cal N}\left( {{\bf{0}},{A_r}{{\sigma^2_r}}{{\bf{R}}}} \right)$ denotes the EMI and ${{\mathbf{n}}_m} \sim \mathcal{C}\mathcal{N}\left( {0,{\sigma ^2}{{\mathbf{I}}_L}} \right)$ denotes the noise. The first part can be written as
\begin{align}
  &\mathbb{E}\left\{ {{\mathbf{H}}_m^H{\mathbf{\Phi n}}{{\mathbf{n}}^H}{{\mathbf{\Phi }}^H}{{\mathbf{H}}_m}} \!\right\} \notag\\
  &= \mathbb{E}\left\{ {{{\left( {{{{\mathbf{\bar H}}}_m} + {{{\mathbf{\tilde H}}}_m}} \right)}^H}{\mathbf{\Phi n}}{{\mathbf{n}}^H}{{\mathbf{\Phi }}^H}\left( {{{{\mathbf{\bar H}}}_m} + {{{\mathbf{\tilde H}}}_m}} \!\right)} \right\} \notag \\
   &= \mathbb{E}\left\{ {{\mathbf{\bar H}}_m^H{\mathbf{\Phi n}}{{\mathbf{n}}^H}{{\mathbf{\Phi }}^H}{{{\mathbf{\bar H}}}_m}} \right\} + \mathbb{E}\left\{ {{\mathbf{\tilde H}}_m^H{\mathbf{\Phi n}}{{\mathbf{n}}^H}{{\mathbf{\Phi }}^H}{{{\mathbf{\tilde H}}}_m}} \right\} \notag \\
   &= \sigma _r^2{A_r}{\mathbf{\bar H}}_m^H{\mathbf{\Phi R}}{{\mathbf{\Phi }}^H}{{{\mathbf{\bar H}}}_m} + \underbrace {\mathbb{E}\left\{ {{\mathbf{\tilde H}}_m^H{\mathbf{\Phi n}}{{\mathbf{n}}^H}{{\mathbf{\Phi }}^H}{{{\mathbf{\tilde H}}}_m}} \right\}}_{{\mathbf{Q}_m}}.
\end{align}
Using the same method as \eqref{Q2mk}, we can obtain
\begin{align}
&{\left[ {{\mathbf{Q}_m}} \right]_{ll'}} = {\left[ {\mathbb{E}\left\{ {{\mathbf{\tilde H}}_m^H{\mathbf{\Phi n}}{{\mathbf{n}}^H}{{\mathbf{\Phi }}^H}{{{\mathbf{\tilde H}}}_m}} \right\}} \right]_{ll'}} \notag\\
&= \sigma _r^2{A_r}{\text{tr}}\left( {{\mathbf{\Phi R}}{{\mathbf{\Phi }}^H}{{\left[ {{{{\mathbf{\tilde R}}}_m}} \right]}_{\left( {lN - N + 1 \sim lN,l'N - N + 1 \sim l'N} \right)}}} \right).
\end{align}
Then, we can obtain $\mathbb{E}\left\{ {{\mathbf{n}}_{mk}^p{{\left( {{\mathbf{n}}_{mk}^p} \right)}^H}} \right\} = {\tau _p}{{\mathbf{R}}_{mm}} + {\tau _p}{\sigma ^2}{{\mathbf{I}}_L}$, where ${{\mathbf{R}}_{mm}} = \sigma _r^2{A_r}{\mathbf{\bar H}}_m^H{\mathbf{\Phi R}}{{\mathbf{\Phi }}^H}{{{\mathbf{\bar H}}}_m} + {\mathbf{Q}_m}$.
\section{Proof of Theorem 3}
The expectations in \eqref{gamma} are calculated here. We begin with the numerator term as
\begin{align}
&\mathbb{E}\left\{ {{{\bf{u}}_{kk}}} \right\} = \mathbb{E}\left\{ {{{\left[ {{\bf{\hat o}}_{1k}^H{{{\bf{\hat o}}}_{1k}}, \cdots ,{\bf{\hat o}}_{Mk}^H{{{\bf{\hat o}}}_{Mk}}} \right]}^T}} \right\}\notag\\
& \!=\! {\left[ {{\rm{tr}}\left( {{\hat p_k}{\tau _p}{{\bf{\Omega }}_{1k}}} \right) \!+\! {{\left\| {{{{\bf{\bar o}}}_{1k}}} \right\|}^2}, \cdots \!,{\rm{tr}}\left( {{\hat p_k}{\tau _p}{{\bf{\Omega }}_{Mk}}} \right) \!+\! {{\left\| {{{{\bf{\bar o}}}_{Mk}}} \right\|}^2}} \right]^T}\!\!.
\end{align}
Similarly, we compute the noise term as
\begin{align}
{\bf{a}}_k^H{{\bf{D}}_k}{{\bf{a}}_k} &= {\bf{a}}_k^H{\rm{diag}}\left( {\mathbb{E}\left\{ {{{\left\| {{{{\bf{\hat o}}}_{1k}}} \right\|}^2}} \right\}, \cdots ,\mathbb{E}\left\{ {{{\left\| {{{{\bf{\hat o}}}_{Mk}}} \right\|}^2}} \right\}} \right){{\bf{a}}_k}\notag\\
& = {\bf{A}}_k^H{\rm{diag}}\left( {{z_{1k}}, \cdots ,{z_{Mk}}} \right){{\bf{A}}_k},
\end{align}
where ${z_{mk}} = {\text{tr}}\left( {{{\hat p}_k}{\tau _p}{{\mathbf{\Omega }}_{mk}} + {{\left\| {{{{\mathbf{\bar o}}}_{mk}}} \right\|}^2}} \right)$.

The interference term expectation in the denominator of \eqref{gamma} is
\begin{align}
&\mathbb{E}\left\{ {{{\left| {\sum\limits_{m = 1}^M {a_{mk}^{}{\bf{\hat o}}_{mk}^H{{\bf{o}}_{mi}}} } \right|}^2}} \right\} = \sum\limits_{m = 1}^M {\sum\limits_{n = 1}^M {{a_{mk}}a_{nk}^ * } } \mathbb{E}\left\{ {{{\left( {{\bf{\hat o}}_{mk}^H{{\bf{o}}_{mi}}} \right)}^H}\left( {{\bf{\hat o}}_{nk}^H{{\bf{o}}_{ni}}} \right)} \right\},
\end{align}
where $\mathbb{E}\{ {{{( {{\bf{\hat o}}_{mk}^H{{\bf{o}}_{mi}}} )}^H}( {{\bf{\hat o}}_{nk}^H{{\bf{o}}_{ni}}} )} \}$ is computed for all the possible APs and UEs combinations. We utilize the independence of channel estimation at different APs. When $m \ne n,i \notin {{\cal P}_k}$, we obtain $\mathbb{E}\left\{ {{{\left( {{\bf{\hat o}}_{mk}^H{{\bf{o}}_{mi}}} \right)}^H}\left( {{\bf{\hat o}}_{nk}^H{{\bf{o}}_{ni}}} \right)} \right\} = 0$. For $m \ne n,i \in {{\cal P}_k}\backslash \left\{ k \right\}$, we derive $\mathbb{E}\left\{ {{{\left( {{\bf{\hat o}}_{mk}^H{{\bf{o}}_{mi}}} \right)}^H}\left( {{\bf{\hat o}}_{nk}^H{{\bf{o}}_{ni}}} \right)} \right\} = \mathbb{E}\left\{ {{\bf{\hat o}}_{mi}^H{{{\bf{\hat o}}}_{mk}}} \right\}\left\{ {{\bf{\hat o}}_{nk}^H{{{\bf{\hat o}}}_{ni}}} \right\}$, where
\begin{align}
\mathbb{E}\left\{ {{\bf{\hat o}}_{mi}^H{{{\bf{\hat o}}}_{mk}}} \right\} = \sqrt {{{\hat p}_k}{{\hat p}_i}} {\tau _p}{\rm{tr}}\left( {{\bf{R}}_{mk}^o{\bf{\Psi }}_{mk}^{ - 1}{\bf{R}}_{mi}^o} \right),
\end{align}
since $\mathbb{E}\left\{ {{{{\bf{\bar o}}}^{H}_{mk}}{{{\bf{\bar o}}}_{mi}}{e^{ - j{\varphi _{mk}}}}{e^{j{\varphi _{mi}}}}} \right\} = 0$ and $\mathbb{E}\left\{ {\sqrt {{\hat p_k}{\tau _p}} \left({\bf{R}}_{mk}^o{\bf{\Psi }}_{mk}^{ - 1}\left( {{\bf{y}}_{mk}^p - {\bf{\bar y}}_{mk}^p} \right)\right)^{H}{{{\bf{\bar o}}}_{mi}}{e^{j{\varphi _{mi}}}}} \right\} = 0$. We repeat the same calculation for AP $n$ and obtain
\begin{align}
&\mathbb{E}\left\{ {{{\left( {{\bf{\hat o}}_{mk}^H{{\bf{o}}_{mi}}} \right)}^H}\left( {{\bf{\hat o}}_{nk}^H{{\bf{o}}_{ni}}} \right)} \right\} = {{\hat p}_k}{{\hat p}_i}\tau _p^2{\rm{tr}}\left( {{\bf{R}}_{mk}^o{\bf{\Psi }}_{mk}^{ - 1}{\bf{R}}_{mi}^o} \right){\rm{tr}}\left( {{\bf{R}}_{ni}^o{\bf{\Psi }}_{nk}^{ - 1}{\bf{R}}_{nk}^o} \right).
\end{align}
For another case $m \ne n,i = k$, we obtain
\begin{align}
&\mathbb{E}\left\{ {{{\left( {{\bf{\hat o}}_{mk}^H{{\bf{o}}_{mi}}} \right)}^H}\left( {{\bf{\hat o}}_{nk}^H{{\bf{o}}_{ni}}} \right)} \right\} = \hat p_k^2\tau _p^2{\rm{tr}}\left( {{{\bf{\Omega }}_{mk}}} \right){\rm{tr}}\left( {{{\bf{\Omega }}_{nk}}} \right) + {\rm{tr}}\left( {{{{\bf{\bar o}}}_{mk}}{\bf{\bar o}}_{mk}^H} \right){\rm{tr}}\left( {{{{\bf{\bar o}}}_{nk}}{\bf{\bar o}}_{nk}^H} \right)\notag\\
&+ {{\hat p}_k}{\tau _p}{\rm{tr}}\left( {{{\bf{\Omega }}_{nk}}} \right){\rm{tr}}\left( {{{{\bf{\bar o}}}_{mk}}{\bf{\bar o}}_{mk}^H} \right) + {{\hat p}_k}{\tau _p}{\rm{tr}}\left( {{{\bf{\Omega }}_{mk}}} \right){\rm{tr}}\left( {{{{\bf{\bar o}}}_{nk}}{\bf{\bar o}}_{nk}^H} \right).
\end{align}
Similarly for $m = n,i = k$, we adopt the same method as in \cite{ozdogan2019performance} and calculate the following equations
\begin{align}
&\mathbb{E}\left\{ {{\mathbf{\tilde o}}_{mk}^H{{{\mathbf{\hat o}}}_{mk}}{\mathbf{\hat o}}_{mk}^H{{{\mathbf{\tilde o}}}_{mk}}} \right\} = {\text{tr}}\!\left( {\left( {{{\hat p}_k}{\tau _p}{{\mathbf{\Omega }}_{mk}} + {{{\mathbf{\bar o}}}_{mk}}{\mathbf{\bar o}}_{mk}^H} \right){{\mathbf{C}}_{mk}}} \right), \\
  &\mathbb{E}\left\{ {{\mathbf{\hat o}}_{mk}^H{{{\mathbf{\hat o}}}_{mk}}{\mathbf{\hat o}}_{mk}^H{{{\mathbf{\hat o}}}_{mk}}} \right\} = {\left( {{{\hat p}_k}{\tau _p}} \right)^2}{\text{tr}}\left( {{{\mathbf{\Omega }}_{mk}}} \right){\text{tr}}\left( {{{\mathbf{\Omega }}_{mk}}} \right) \notag\\
  &+ {{\hat p}_k}{\tau _p}{\text{tr}}\left( {\left( {{\mathbf{R}}_{mk}^o - {{\mathbf{C}}_{mk}}} \right){{\mathbf{\Omega }}_{mk}}} \right) + {{\hat p}_k}{\tau _p}{\mathbf{\bar o}}_{mk}^H{{\mathbf{\Omega }}_{mk}}{{{\mathbf{\bar o}}}_{mk}} \notag\\
  &+ {\mathbf{\bar o}}_{mk}^H\left( {{\mathbf{R}}_{mk}^o - {{\mathbf{C}}_{mk}}} \right){{{\mathbf{\bar o}}}_{mk}} + {\mathbf{\bar o}}_{mk}^H{{{\mathbf{\bar o}}}_{mk}}{\mathbf{\bar o}}_{mk}^H{{{\mathbf{\bar o}}}_{mk}} + 2{{\hat p}_k}{\tau _p}{\mathbf{\bar o}}_{mk}^H{{{\mathbf{\bar o}}}_{mk}}{\text{tr}}\left( {{{\mathbf{\Omega }}_{mk}}} \right).
\end{align}
Combining the above two formulas yields the result for $m = n,i = k$ as
\begin{align}
  \mathbb{E}\left\{ {{{\left| {{\mathbf{\hat o}}_{mk}^H{{\mathbf{o}}_{mk}}} \right|}^2}} \right\}
   &= \hat p_k^2\tau _p^2{\left| {{\text{tr}}\left( {{{\mathbf{\Omega }}_{mk}}} \right)} \right|^2} + {{\hat p}_k}{\tau _p}{\text{tr}}\left( {{{\mathbf{\Omega }}_{mk}}{\mathbf{R}}_{mk}^o} \right) \notag\\
   &+ {\mathbf{\bar o}}_{mk}^H{\mathbf{R}}_{mk}^o{{{\mathbf{\bar o}}}_{mk}} + {{\hat p}_k}{\tau _p}{\mathbf{\bar o}}_{mk}^H{{\mathbf{\Omega }}_{mk}}{{{\mathbf{\bar o}}}_{mk}} \notag \\
   &+ 2{{\hat p}_k}{\tau _p}{\text{tr}}\left( {{{\mathbf{\Omega }}_{mk}}} \right){\mathbf{\bar o}}_{mk}^H{{{\mathbf{\bar o}}}_{mk}} + {\text{tr}}{\left( {{{{\mathbf{\bar o}}}_{mk}}{\mathbf{\bar o}}_{mk}^H} \right)^2}.
\end{align}
Then, for the case $m = n,i \notin {{\cal P}_k}$, we obtain
\begin{align}
\mathbb{E}\left\{ {{{\left| {{\bf{\hat o}}_{mk}^H{{\bf{o}}_{mi}}} \right|}^2}} \right\} &= {\hat p_k}{\tau _p}{\rm{tr}}\left( {{\bf{R}}_{mi}^o{{\bf{\Omega }}_{mk}}} \right) + {\bf{\bar o}}_{mk}^H{\bf{R}}_{mi}^o{{{\bf{\bar o}}}_{mk}} + {\hat p_k}{\tau _p}{\bf{\bar o}}_{mi}^H{{\bf{\Omega }}_{mk}}{{{\bf{\bar o}}}_{mi}} + {\left| {{\bf{\bar o}}_{mk}^H{{{\bf{\bar o}}}_{mi}}} \right|^2}.
\end{align}
For $m = n,i \in {{\cal P}_k}\backslash \left\{ k \right\}$, we obtain
\begin{align}
\mathbb{E}\left\{ {{{\left| {{\bf{\hat o}}_{mk}^H{{\bf{o}}_{mi}}} \right|}^2}} \right\} &= {{\hat p}_k}{\tau _p}{\rm{tr}}\left( {{{\bf{\Omega }}_{mk}}{\bf{R}}_{mi}^o} \right){\rm{ + tr}}\left( {{{{\bf{\bar o}}}_{mk}}{\bf{\bar o}}_{mk}^H{\bf{R}}_{mi}^o} \right) \notag \\
\!&+\! {\left| {{\bf{\bar o}}_{mk}^H{{{\bf{\bar o}}}_{mi}}} \right|^2} \!+\! {{\hat p}_k}{{\hat p}_i}\tau _p^2{\left| {{\rm{tr}}\left( {{\bf{R}}_{mk}^o{\bf{\Psi }}_{mk}^{ - 1}{\bf{R}}_{mk}^o} \right)} \right|^2}\!.
\end{align}
Finally, arranging all these equations yields \eqref{65} at the top of this page.

\begin{figure*}[t!]
\normalsize
\setcounter{mytempeqncnt}{1}
\setcounter{equation}{64}
\begin{align}\label{65}
  &\mathbb{E}\left\{ {{{\left| {\sum\limits_{m = 1}^M {a_{mk}^ * {\mathbf{v}}_{mk}^H{{\mathbf{o}}_{mi}}} } \right|}^2}} \right\} \notag\\
  &= \sum\limits_{m = 1}^M {{{\left| {{a_{mk}}} \right|}^2}\left[ {{p_k}{\tau _p}{\text{tr}}\left( {{\mathbf{R}}_{mi}^o{{\mathbf{\Omega }}_{mk}}} \right) + {\mathbf{\bar o}}_{mk}^H{\mathbf{R}}_{mi}^o{{{\mathbf{\bar o}}}_{mk}} + {p_k}{\tau _p}{\mathbf{\bar o}}_{mi}^H{{\mathbf{\Omega }}_{mk}}{{{\mathbf{\bar o}}}_{mi}} + {{\left| {{\mathbf{\bar o}}_{mk}^H{{{\mathbf{\bar o}}}_{mi}}} \right|}^2}} \right]} \notag\\
   &+ \left\{ {\begin{array}{*{20}{c}}
  {\sum\limits_{m = 1}^M {{{\left| {{a_{mk}}} \right|}^2}{{\hat p}_k}{{\hat p}_i}\tau _p^2{{\left| {{\text{tr}}\left( {{\mathbf{R}}_{mi}^o{\mathbf{\Psi }}_{mk}^{ - 1}{\mathbf{R}}_{mk}^o} \right)} \right|}^2},} }&{i \in {\mathcal{P}_k}\backslash \left\{ k \right\}}, \notag \\
  {\sum\limits_{m = 1}^M {{{\left| {{a_{mk}}} \right|}^2}\left[ {\hat p_k^2\tau _p^2{{\left| {{\text{tr}}\left( {{{\mathbf{\Omega }}_{mk}}} \right)} \right|}^2} + 2\sqrt {{{\hat p}_k}{{\hat p}_i}} {\tau _p}{\text{tr}}\left( {{{\mathbf{\Omega }}_{mk}}} \right){\mathbf{\bar o}}_{mk}^H{{{\mathbf{\bar o}}}_{mk}}} \right],} }&{i = k}, \notag \\
  {0,}&{i \notin {\mathcal{P}_k}},
\end{array}} \right. \notag \\
   &+ \left\{ {\begin{array}{*{20}{c}}
  {\sum\limits_{m = 1}^M {\sum\limits_{n = 1}^M {{a_{mk}}a_{nk}^ * \left[ {{{\hat p}_k}{{\hat p}_i}\tau _p^2{\text{tr}}\left( {{\mathbf{R}}_{mk}^o{\mathbf{\Psi }}_{mk}^{ - 1}{\mathbf{R}}_{mi}^o} \right){\text{tr}}\left( {{\mathbf{R}}_{nk}^o{\mathbf{\Psi }}_{nk}^{ - 1}{\mathbf{R}}_{ni}^o} \right) + {{\hat p}_k}{\tau _p}{\text{tr}}\left( {{{\mathbf{\Omega }}_{nk}}} \right){\text{tr}}\left( {{{{\mathbf{\bar o}}}_{mk}}{\mathbf{\bar o}}_{mk}^H} \right)} \right.} } }&{} \\
  {\left. { + {{\hat p}_k}{\tau _p}{\text{tr}}\left( {{{\mathbf{\Omega }}_{mk}}} \right){\text{tr}}\left( {{{{\mathbf{\bar o}}}_{nk}}{\mathbf{\bar o}}_{nk}^H} \right) + {\text{tr}}\left( {{{{\mathbf{\bar o}}}_{mk}}{\mathbf{\bar o}}_{mk}^H} \right){\text{tr}}\left( {{{{\mathbf{\bar o}}}_{nk}}{\mathbf{\bar o}}_{nk}^H} \right)} \right],}&\!\!\!\!\!\!\!\!\!\!\!\!\!\!\!\!\!\!\!{i \in {\mathcal{P}_k}}, \\
  {0,}&\!\!\!\!\!\!\!\!\!\!\!\!\!\!\!\!\!\!\!{i \notin {\mathcal{P}_k}}.
\end{array}} \right.
\end{align}
\setcounter{equation}{65}
\hrulefill
\end{figure*}

Then, for the EMI term ${{\mathbf{U}}_k} = {\text{diag}}( {\mathbb{E}\{ {{{\| {{\mathbf{\hat o}}_{1k}^H{\mathbf{H}}_1^H{\mathbf{\Phi n}}} \|}^2}} \}, \cdots ,\mathbb{E}\{ {{{\| {{\mathbf{\hat o}}_{Mk}^H{\mathbf{H}}_M^H{\mathbf{\Phi n}}} \|}^2}} \}} )$, we calculate as follow
\begin{align}\label{66}
  &\mathbb{E}\left\{ {{{\left\| {{\mathbf{\hat o}}_{mk}^H{\mathbf{H}}_m^H{\mathbf{\Phi n}}} \right\|}^2}} \right\} = \mathbb{E}\left\{ {{\mathbf{\hat o}}_{mk}^H{\mathbf{H}}_m^H{\mathbf{\Phi n}}{{\mathbf{n}}^H}{{\mathbf{\Phi }}^H}{{\mathbf{H}}_m}{{{\mathbf{\hat o}}}_{mk}}} \right\} \notag \\
   &= \sigma _r^2{A_r}\mathbb{E}\left\{ {{{\left( {{{{\mathbf{\bar o}}}_{mk}}{e^{j{\varphi _k}}} + \sqrt {{{\hat p}_k}} {\mathbf{R}}_{mk}^o{\mathbf{\Psi }}_{mk}^{ - 1}\left( {{\mathbf{y}}_{mk}^p - {\mathbf{\bar y}}_{mk}^p} \right)} \right)}^H}{\mathbf{H}}_m^H{\mathbf{\Phi R}}{{\mathbf{\Phi }}^H}{{\mathbf{H}}_m}} \right. \notag \\
   &\times \left. {\left( {{{{\mathbf{\bar o}}}_{mk}}{e^{j{\varphi _k}}} + \sqrt {{{\hat p}_k}} {\mathbf{R}}_{mk}^o{\mathbf{\Psi }}_{mk}^{ - 1}\left( {{\mathbf{y}}_{mk}^p - {\mathbf{\bar y}}_{mk}^p} \right)} \right)} \right\} = \sigma _r^2{A_r}\mathbb{E}\left\{ {{{{\mathbf{\bar o}}}^H}_{mk}{\mathbf{H}}_m^H{\mathbf{\Phi R}}{{\mathbf{\Phi }}^H}{{\mathbf{H}}_m}{{{\mathbf{\bar o}}}_{mk}}} \right\} \notag \\
   &+ \sigma _r^2{A_r}\mathbb{E}\left\{ {\sqrt {{{\hat p}_k}} {{\left( {{\mathbf{y}}_{mk}^p - {\mathbf{\bar y}}_{mk}^p} \right)}^H}{{\left( {{\mathbf{\Psi }}_{mk}^{ - 1}} \right)}^H}{{\left( {{\mathbf{R}}_{mk}^o} \right)}^H}{\mathbf{H}}_m^H{\mathbf{\Phi R}}{{\mathbf{\Phi }}^H}{{\mathbf{H}}_m}\sqrt {{{\hat p}_k}} {\mathbf{R}}_{mk}^o{\mathbf{\Psi }}_{mk}^{ - 1}\left( {{\mathbf{y}}_{mk}^p - {\mathbf{\bar y}}_{mk}^p} \right)} \right\}.
\end{align}
Then first term can be obtain by Appendix B as
\begin{align}
  \mathbb{E}\!\left\{ {{{{\mathbf{\bar o}}}^H}_{mk}{\mathbf{H}}_m^H{\mathbf{\Phi R}}{{\mathbf{\Phi }}^H}{{\mathbf{H}}_m}{{{\mathbf{\bar o}}}_{mk}}} \right\} \!\!&=\! \sigma _r^2{A_r}{{{\mathbf{\bar o}}}^H}_{mk}{\mathbf{\bar H}}_m^H{\mathbf{\Phi R}}{{\mathbf{\Phi }}^H}{{{\mathbf{\bar H}}}_m}{{{\mathbf{\bar o}}}_{mk}} \notag \\
   &\!+\! \sigma _r^2{A_r}{{{\mathbf{\bar o}}}^H}_{mk}\!{\left[ \!{{\text{tr}}\left(\!\! {{\mathbf{\Phi R}}{{\mathbf{\Phi }}^H}\!{{\left[ {{{{\mathbf{\tilde R}}}_m}} \!\right]}_{\left(\! {lN \!-\! N \!+\! 1 \sim lN,l'N \!-\! N \!+\! 1 \sim l'N} \!\right)}}} \!\right)} \!\right]_{\!ll'}}\!\!{{{\mathbf{\bar o}}}_{mk}}.
\end{align}

The second term is derived as
\begin{align}\label{68}
  &\mathbb{E}\left\{ {\sqrt {{{\hat p}_k}} {{\left( {{\mathbf{y}}_{mk}^p - {\mathbf{\bar y}}_{mk}^p} \right)}^H}{{\left( {{\mathbf{\Psi }}_{mk}^{ - 1}} \right)}^H}{{\left( {{\mathbf{R}}_{mk}^o} \right)}^H}{\mathbf{H}}_m^H{\mathbf{\Phi R}}{{\mathbf{\Phi }}^H}{{\mathbf{H}}_m}\sqrt {{{\hat p}_k}} {\mathbf{R}}_{mk}^o{\mathbf{\Psi }}_{mk}^{ - 1}\left( {{\mathbf{y}}_{mk}^p - {\mathbf{\bar y}}_{mk}^p} \right)} \right\} \notag \\
   &= {{\hat p}_k}\tau _p^2{{\mathbf{\Psi }}_{mk}}{\left( {{\mathbf{\Psi }}_{mk}^{ - 1}} \right)^H}{\left( {{\mathbf{R}}_{mk}^o} \right)^H}{\mathbf{\bar H}}_m^H{\mathbf{\Phi R}}{{\mathbf{\Phi }}^H}{{{\mathbf{\bar H}}}_m}{\mathbf{R}}_{mk}^o{\mathbf{\Psi }}_{mk}^{ - 1} \notag \\
   &+ {{\hat p}_k}\mathbb{E}\left\{ {{{\left( {{\mathbf{y}}_{mk}^p - {\mathbf{\bar y}}_{mk}^p} \right)}^H}{{\left( {{\mathbf{\Psi }}_{mk}^{ - 1}} \right)}^H}{{\left( {{\mathbf{R}}_{mk}^o} \right)}^H}{\mathbf{H}}_m^H{\mathbf{\Phi R}}{{\mathbf{\Phi }}^H}{{\mathbf{H}}_m}{\mathbf{R}}_{mk}^o{\mathbf{\Psi }}_{mk}^{ - 1}\left( {{\mathbf{y}}_{mk}^p - {\mathbf{\bar y}}_{mk}^p} \right)} \right\} \notag \\
   &= {{\hat p}_k}\tau _p^2{\text{tr}}\left( {{{\left( {{\mathbf{\Psi }}_{mk}^{ - 1}} \right)}^H}{{\left( {{\mathbf{R}}_{mk}^o} \right)}^H}{\mathbf{\bar H}}_m^H{\mathbf{\Phi R}}{{\mathbf{\Phi }}^H}{{{\mathbf{\bar H}}}_m}{\mathbf{R}}_{mk}^o} \right) \notag \\
   &+ {{\hat p}_k}\tau _p^2{\text{tr}}\left( {{{\left( {{\mathbf{\Psi }}_{mk}^{ - 1}} \right)}^H}{{\left( {{\mathbf{R}}_{mk}^o} \right)}^H}{{\left[ {{\text{tr}}\left( {{\mathbf{\Phi R}}{{\mathbf{\Phi }}^H}{{\left[ {{{{\mathbf{\tilde R}}}_m}} \right]}_{\left( {lN - N + 1 \sim lN,l'N - N + 1 \sim l'N} \right)}}} \right)} \right]}_{ll'}}{\mathbf{R}}_{mk}^o} \right).
\end{align}
By combining the first and second terms, we can derive the expectation of EMI as
\begin{align}
&\mathbb{E}\left\{ {{{\left\| {{\mathbf{\hat o}}_{mk}^H{\mathbf{H}}_m^H{\mathbf{\Phi n}}} \right\|}^2}} \right\} = {{{\mathbf{\bar o}}}^H}_{mk}{{\mathbf{R}}_{mn}}{{{\mathbf{\bar o}}}_{mk}} + {{\hat p}_k}\tau _p^2{\text{tr}}\left( {{{\left( {{\mathbf{\Psi }}_{mk}^{ - 1}} \right)}^H}{{\left( {{\mathbf{R}}_{mk}^o} \right)}^H}{{\mathbf{R}}_{mn}}{\mathbf{R}}_{mk}^o} \right).
\end{align}
Finally, combining the above cases, we can derive the expectation of \eqref{gamma}.

\end{appendices}

\bibliographystyle{IEEEtran}
\bibliography{IEEEabrv,RIS_aided_EMI}

\end{document}